\documentstyle[prl,aps,epsfig,floats]{revtex}

\newcommand{\bib}{\bibitem}
\newcommand\bea{\begin{eqnarray}}
\newcommand\eea{\end{eqnarray}}
\newcommand\beq{\begin{equation}}
\newcommand\eeq{\end{equation}}
\newcommand\non{\nonumber}
\newcommand{\ua}{\uparrow}
\newcommand{\da}{\downarrow}

\begin{document}

\draft

\textheight=24cm
\twocolumn[\hsize\textwidth\columnwidth\hsize\csname@twocolumnfalse\endcsname

\title{\Large \bf Conductance of quantum wires: a numerical study of the 
effects of an impurity and interactions}
\author{\bf Amit Agarwal and Diptiman Sen}
\address{\it Centre for High Energy Physics, Indian Institute of Science,
Bangalore 560012, India}

\date{\today}
\maketitle

\begin{abstract}
We use the non-equilibrium Green's function formalism and a 
self-consistent Hartree-Fock approximation to numerically study the effects 
of a single impurity and interactions between the electrons (with and without 
spin) on the conductance of a quantum wire. We study how the conductance varies
with the wire length, the temperature, and the strengths of the impurity and 
interactions. The numerical results for the dependence of the conductance on 
the wire length and temperature are compared with the results obtained from a
renormalization group analysis based on the Hartree-Fock approximation. For the
spin-1/2 model with a repulsive on-site interaction or the spinless model with
an attractive nearest neighbor interaction, we find that the conductance 
increases with increasing wire length or decreasing temperature. This can be 
explained using the Born approximation in scattering theory. For a strong 
impurity, the conductance is significantly different for a repulsive and an 
attractive impurity; this is due to the existence of a bound state in the 
latter case. In general, the large density deviations close to the impurity
have an appreciable effect on the conductance at short distances which is 
not captured by the renormalization group equations.
\end{abstract}
\vskip .5 true cm

\vskip.5pc
]

\section{Introduction}

The conductance of electrons in a quantum wire has been the subject of 
intensive study in recent years, both experimentally 
\cite{tarucha,liang,bkane,yacoby,aus,reilly} and theoretically
\cite{kane,furusaki,lal1}. For a wire in which only one channel is 
available to the electrons and the transport is 
ballistic (i.e., there are no impurities inside the wire, and there is no 
scattering from phonons or from the contacts between the wire and its two 
leads), the conductance is given by $G = 2e^2 /h$ for infinitesimal bias 
\cite{datta,imry}. This result is expected to hold even if one takes into 
account the interactions between the electrons since such two-body scatterings
conserve the momentum. However, if there is an impurity inside the wire which 
scatters the electrons, then the conductance is reduced 
because such a scattering does not conserve the momentum of the electron.
For a one-dimensional system containing a $\delta$-function impurity
with strength $V$, we obtain
\beq
G ~=~ \frac{2e^2}{h} ~ (1 ~-~ c ~V^2 ~) ~,
\eeq
to lowest order in $V$, where $c$ is a constant related to the Fermi velocity
of the electrons (see the discussion below Eq. (\ref{rt1})). (The situation 
is different if the wire is only quasi-one-dimensional, and the impurity 
potential has a finite range \cite{chu}). In the absence of interactions, 
$G$ does not depend on the wire length $L$ or the 
temperature $T$ (as long as $k_B T$ is much less than the Fermi energy). 
But in the presence of interactions, it turns out that $V$ effectively 
becomes a function of the length scale (which is related to either $L$ 
or $T$ as will be explained below), and $G$ therefore varies with $L$ and 
$T$. The variation of $V$ with length scale is governed by a renormalization
group (RG) equation. (In this paper, we will only consider a non-magnetic
impurity which scatters electrons in a spin independent way).

There are three length scales which are of interest in the problem. The 
smallest of them is $\lambda \equiv \pi /k_F$ (where $k_F$ is the Fermi 
wavenumber); this is the wavelength of the oscillations in the electronic 
density near impurities as we will see. The other two length scales are the 
length of the wire $L$, and the thermal coherence length $L_T$ which is equal 
to $\hbar v_F /(k_B T)$ (where $v_F$ is the Fermi velocity). $L_T$ gives an
idea of the distance beyond which an electron wave function loses its
phase coherence. At a temperature $T$, the conductance typically receives 
contributions from a number of states near the Fermi energy whose energies
have a spread of the order of $\Delta E = k_B T$. Hence the spread in 
momentum $\Delta p = \Delta E /v_F$ is of the order of $k_B T /v_F$. 
A superposition of waves with such a spread of momenta loses phase coherence 
in a distance of the order of $L_T =\hbar v_F /(k_B T)$. Electronic transport
is therefore thermally incoherent if the wire length $L >> L_T$, coherent if 
$L << L_T$, and partially coherent in the intermediate range. This will
become clearer when we discuss our numerical results.

The RG equations for the transmission coefficient $|t|^2$ has been derived 
from continuum theories in several ways \cite{kane,furusaki,yue,lal2,polyakov}.
We will discuss these equations for two models in Sec. II. (The conductance 
is related to the transmission as $G = (2e^2 /h) |t|^2$, where the factor of 
2 is due to the electron spin). The RG equation is used to analytically
follow the evolution of $|t|^2$ starting from a short distance scale 
and going up to a length scale which is the smaller of 
the two quantities $L$ and $L_T$. If $L >> L_T$, the conductance is
governed by $L_T$ and not $L$, and vice versa if $L_T >> L$. The RG equation 
has two fixed points which lie at $|t|^2 =0$ and 1; the system
approaches one of these fixed points if $L$ and $L_T$ are both very large.

The above statements are only valid for length scales much longer than 
$\lambda$ because only then can one use a continuum description from which 
the RG equation is derived. It would therefore be useful to consider an 
alternative method for computing the conductance which works even at short 
length scales where the continuum description is not valid. An interesting 
thing which occurs at short distances is that if the impurity provides an 
attractive potential to the electrons, there is a bound state whose
wave function decays exponentially away from the impurity. We may wonder what 
effect a bound state has on the conductance; the RG equations mentioned above
do not take this state into account. One may think that a bound state (whose 
energy lies outside the bandwidth of the leads) cannot directly affect the 
conductance, because electrons coming in from or going out to the leads 
cannot enter or leave such states in the absence of any inelastic scattering.
However, such a state contributes to the electronic density near the impurity,
and that can affect the transport in the presence of interactions. 

In a series of papers, the technique of functional RG has been used to 
numerically study the conductance and other properties of interacting electron
systems in one dimension \cite{enss,ander,barnabe,meden}. The authors of those
papers have gone up to very large system sizes and have found excellent 
agreement at those length scales between their numerical results and the 
asymptotic scaling forms given by the RG equations.

In this paper, we will use the non-equilibrium Green's function (NEGF) 
formalism to numerically study the conductance of a quantum wire 
\cite{datta,meir,datta2,dhar,tsukada}. Since we will use lattice models 
and the Hartree-Fock (HF) approximation for dealing with interactions
in our numerical studies, we will first discuss those topics in Sec. III.
In that section, we will also show how the Born approximation for scattering
can qualitatively explain the dependence of the conductance on the wire
length which we obtain from the RG equations in Sec. II.

The NEGF formalism will be briefly described in Sec. IV. The advantage of 
this method is that it treats the infinitely extended leads (reservoirs) in 
an exact way, and it can be used for all values of the wire length and the 
impurity strength. However, it is accurate only for weak interactions between 
the electrons because we are forced to use a HF approximation for 
dealing with the interactions (for reasons which will be explained below). We 
will use a lattice model for both the wire and its leads. The 
Hamiltonian will have hopping terms in the wire and in the 
leads, and a density-density interaction (on-site for spin-1/2 electrons, and
between nearest neighbor sites for spinless electrons) only in the wire.

In Sec. V, we will describe our results for spin-1/2 and spinless electrons 
for different values of the impurity potential and the interaction strength, 
and we will compare our results with those obtained by the RG analysis. For 
the case of spinless electrons, we find that the agreement between the RG
and numerical results is excellent if we follow a certain procedure. In 
Sec. VI, we will make some concluding remarks.

\section{Renormalization group equation for scattering from a point}

There are several ways of studying the renormalization group (RG) evolution 
of the scattering from one or more points in a one-dimensional system of 
interacting electrons. One can use the technique of bosonization 
\cite{kane,furusaki,gogolin}, a fermionic RG method \cite{yue,lal2,polyakov},
and the functional RG method \cite{enss,ander,barnabe,meden}. Since our 
numerical calculations use a HF approximation, the method of Refs. 
\cite{yue,lal2} will be the most useful for us. 
Before considering the HF approach, however, we will briefly discuss the 
RG equations obtained by bosonization for spinless electrons.

\subsection{Spinless electrons}

Let us first consider the Hamiltonian for non-interacting electrons in
the presence of a $\delta$-function impurity placed at the origin,
\beq
H ~=~ \frac{{\vec p}^2}{2m} ~+~ V \delta (x) ~.
\label{ham}
\eeq
It is easy to check that for plane waves incident either from the left or
from the right with wavenumber $k$, the reflection and transmission 
amplitudes are given by \cite{merzbacher}
\bea
r(k) &=& - ~\frac{imV}{\hbar^2 k ~+~ imV} ~, \non \\
{\rm and} \quad t(k) &=& \frac{\hbar^2 k}{\hbar^2 k ~+~ imV} ~.
\label{rt1}
\eea
If the Fermi energy of the electrons is given by $E_F = \hbar^2 k_F^2 /(2m)$ 
(where $k_F$ is the Fermi wavenumber, and $v_F = \hbar k_F /m$ is the Fermi 
velocity), then the conductance $G$ for spinless electrons at zero temperature
is given by $e^2 /h$ times $|t (k_F)|^2$. For $|V| << \hbar v_F$, we see that
$|t (k_F)|^2 = 1 - (V/\hbar v_F)^2$ up to order $V^2$. (We will usually set 
Planck's constant $\hbar = 1$). Eq. (\ref{ham}) has a bound state if
$V < 0$, but this does not play any role in the RG analysis described below.

Now let us introduce interactions between the electrons. We assume a 
density-density interaction between spinless electrons of
the form 
\beq
H_{\rm int} ~=~ \frac{1}{2} ~\int \int ~dx dy ~\rho (x) ~U(x-y) ~\rho (y) ~,
\label{hint}
\eeq
where the density $\rho$ is given in terms of the second-quantized electron 
field $\Psi (x)$ as $\rho = \Psi^\dagger \Psi$. The electron field can be 
written in terms of the right and left moving fields $\Psi_R$ and $\Psi_L$ 
(whose variations in space are governed by wavenumbers much smaller than 
$k_F$) as
\beq
\Psi (x) ~=~ \Psi_R e^{ik_F x} ~+~ \Psi_L e^{-ik_F x} ~.
\label{psirl}
\eeq
If the range of the interaction $U(x)$ is short (of the order of $\lambda$),
such as that of a screened Coulomb repulsion, the Hamiltonian in (\ref{hint})
can be written in the form
\beq
H_{\rm int} ~=~ g_2 ~\int dx ~\Psi_R^\dagger \Psi_R \Psi_L^\dagger \Psi_L ~,
\label{hint2}
\eeq
where $g_2$ is related to the Fourier transform of $U(x)$ as $g_2 = {\tilde U}
(0)- {\tilde U} (2k_F)$. It is convenient to define the dimensionless constant
\beq
\alpha ~=~ \frac{g_2}{2\pi v_F} ~.
\label{alpha}
\eeq

A system of interacting electrons in one dimension such as the one introduced 
above is described by Tomonaga-Luttinger liquid (TLL) theory. The low energy 
excitations of a TLL are particle-hole pairs which are bosonic in nature and 
have a linear relation between energy and momentum. For spinless electrons, 
the low energy and long distance properties of the TLL are governed by three 
quantities, namely, the velocity $v$ of the low energy excitations, a 
dimensionless parameter $K$ which is related to the interaction strength, and 
the Fermi wavenumber $k_F$. For the model described above, we find that 
\cite{gogolin}
\bea
v ~&=&~ v_F ~(1 - \alpha^2 )^{1/2} ~, \non \\
{\rm and} \quad K ~&=&~ \bigl( \frac{1-\alpha}{1+\alpha} \bigr)^{1/2} ~.
\eea
Thus $K=1$ for non-interacting fermions.
For weak interactions, $v=v_F$ and $K = 1 -\alpha$
to first order in $\alpha$. In this paper, we will be interested in the
case in which the interaction is weak, i.e., $|\alpha| << 1$.

It turns out that in the presence of interactions, the impurity strength $V$ 
effectively becomes a function of a length scale $l$, and satisfies a RG
equation. On bosonizing the TLL theory \cite{kane,furusaki,gogolin}, we 
obtain the equation
\beq
\frac{dV}{d \ln l} ~=~(1 ~-~ K) ~ V ~, 
\label{rg1} 
\eeq
to first order in $V$, i.e., this is valid in the weak barrier limit.
In the strong barrier limit, which implies $|V| >> v_F$, bosonization leads
to a different RG equation
\beq
\frac{d (1/V)}{d \ln l} ~=~(1 ~-~ \frac{1}{K}) ~ (1/V) ~, 
\label{rg2} 
\eeq
which is valid to first order in $1/V$. We see that $V=0$ ($V = \infty$)
is a stable fixed point if $K > 1$ ($K < 1$ respectively). Since the above 
RG equations are not valid if $|V|/v_F$ is of order 1, we cannot 
conclude from them whether or not there is an intermediate fixed point. 

For a given value of the wire 
length $L$ and temperature $T$, the RG equations can be used to compute the 
conductance analytically as follows. We begin at a short length scale
(of order $\lambda$) with an initial value of $V$; we then use Eq. (\ref{rg1})
or (\ref{rg2}), depending on whether $V$ is small or large, to follow the 
evolution of $V$ with the length scale $l$. The RG flow stops when $l$ reaches
a distance of the order of $L$ or $L_T$, whichever is {\it smaller}. When the 
RG flow stops, we take the value of $V$ obtained at that point, and compute 
the conductance $G$ in terms of the transmission amplitude given in Eq. 
(\ref{rt1}). [We will see later that the RG flow of the conductance does not 
stop abruptly at one particular length scale. There is an intermediate
range of length scales where the conductance continues to evolve slowly.]

Let us now discuss the second method for obtaining the RG equations, namely,
using a HF approximation for the case of weak interactions
\cite{yue,lal2,polyakov}. This method directly gives an RG equation for
the scattering matrix which is produced by the impurity. The idea is 
that reflection from the impurity leads to an interference between the
incoming and outgoing electron waves; this leads to
Friedel oscillations in the density with an amplitude proportional to $r$
and a wavelength given by $\lambda = \pi /k_F$.
In the presence of interactions, these density oscillations cause the
electrons to scatter; these scatterings therefore renormalize the scattering 
caused by the impurity. The RG equations obtained by this method are given by
\bea
\frac{dt}{d \ln l} &=& - ~\alpha ~t ~|r|^2 ~, \non \\
{\rm and} \quad \frac{dr}{d \ln l} &=& \alpha ~r ~|t|^2 ~, \non \\
\label{rg3}
\eea
to first order in the interaction parameter $\alpha$. (Eq. (\ref{rg3}) is 
consistent with the unitarity of the scattering matrix, namely, $|r|^2 + |t|^2
= 1$ and $r^* t + t^* r = 0$). The solution of Eq. (\ref{rg3}) is given by
\beq
|t (l)|^2 ~=~ \frac{|t (d)|^2 ~(d/l)^{2\alpha}}{1 ~-~ |t(d)|^2 ~+~ |t(d)|^2 ~
(d/l)^{2\alpha}} ~,
\label{solvrg}
\eeq
where $d$ is a short distance scale which is of the order of $\lambda$. 
However, we will see at the end of Sec. V C that it is necessary to choose 
$d$ to be substantially larger than $\lambda$ in order to obtain a good fit 
between the numerical results and the expression in Eq. (\ref{solvrg}).
[At $l \rightarrow \infty$, Eq. (\ref{solvrg}) has fixed points at $|t|=$ 0 
and 1, depending on the sign of $\alpha$; there is no fixed point at 
intermediate values of $t$.]

We can check that Eq. (\ref{rg3}) is equivalent to Eqs. (\ref{rg1}) and 
(\ref{rg2}) if $\alpha$ is small and $|t|$ is close to 1 and 0 respectively. 
If $\alpha$ is small, the Luttinger parameter $K \simeq 1 - \alpha$. 

We note that Eq. (\ref{rg3}) is complementary to Eqs. (\ref{rg1}) and 
(\ref{rg2}) obtained from bosonization in the following sense. Eq. 
(\ref{rg3}) is valid for all values of $t$ but only small values of $\alpha$.
On the other hand, Eqs. (\ref{rg1}) and (\ref{rg2}) are valid for arbitrary 
values of the interaction parameter $\alpha$ but only for
small values of $V$ or $1/V$, i.e., only when either the reflection 
amplitude $r$ or the transmission amplitude $t$ is small. 

\subsection{Spin-$1/2$ electrons}

We now turn to the more realistic case of electrons with spin. For this 
situation, we will only discuss the RG equations obtained from the HF 
approximation for the case of weak interactions \cite{yue,lal2}. We again 
consider a $\delta$-function impurity at one point in the TLL, and a 
density-density interaction of the form given in (\ref{hint}); the density 
is now given by
\beq
\rho ~=~ \Psi_\ua^\dagger \Psi_\ua ~+~ \Psi_\da^\dagger \Psi_\da ~,
\eeq
where $\ua$ and $\da$ denote spin-up and spin-down respectively. Introducing 
right and left moving fields $\Psi_{R,\sigma}$ and $\Psi_{L,\sigma}$ as in 
Eq. (\ref{psirl}), with $\sigma = \ua$ or $\da$, we obtain the interaction
Hamiltonian
\bea
& & H_{\rm int} = \non \\
& & \int dx \sum_{\sigma,\sigma^\prime} [g_1
\Psi_{R\sigma}^\dagger \Psi_{L\sigma^\prime}^\dagger \Psi_{R\sigma^\prime}
\Psi_{L\sigma}+g_2 \Psi_{R\sigma}^\dagger \Psi_{L\sigma^\prime}^\dagger
\Psi_{L\sigma^\prime} \Psi_{R\sigma} \non \\
& & ~~~~~~~~ + \frac{1}{2} g_4 ( \Psi_{R\sigma}^\dagger
\Psi_{R\sigma^\prime}^\dagger \Psi_{R\sigma^\prime} \Psi_{R\sigma} +
\Psi_{L\sigma}^\dagger \Psi_{L\sigma^\prime}^\dagger \Psi_{L\sigma^\prime}
\Psi_{L\sigma} )], \non \\
& &
\eea
where $g_1 = {\tilde U} (2k_F)$, and $g_2 = g_4 = {\tilde U} (0)$. (We 
have ignored umklapp scattering terms here; they only arise if the model
is defined on a lattice and we are at half-filling). It is known that 
$g_1$, $g_2$ and $g_4$ satisfy some RG equations \cite{solyom}; 
the solutions of the lowest order RG equations are given by \cite{yue}
\bea
g_1 (l) &=& \frac{{\tilde U} (2k_F)}{1 ~+~ \frac{{\tilde U} (2k_F)}{\pi v_F}
\ln l} ~, \non \\
g_2 (l) &=& {\tilde U} (0) ~-~ \frac{1}{2} ~{\tilde U} (2k_F) ~+~ \frac{1}{2}~
\frac{{\tilde U} (2k_F)}{1 ~+~ \frac{{\tilde U} (2k_F)}{\pi v_F} \ln l} ~, 
\non \\
g_4 (l) &=& {\tilde U} (0) ~.
\label{g124}
\eea

Next, we can use the HF approximation and the existence of Friedel
oscillations in the density to derive the RG equation of the transmission
amplitude $t$. We find that \cite{yue,lal2}
\beq
\frac{dt}{d \ln l} ~=~ -~ \frac{g_2 (l) ~-~ 2 g_1 (l)}{2 \pi v_F} ~t ~|r|^2 ~.
\label{rg4}
\eeq
Due to the RG flow of the couplings $g_1$ and $g_2$, the solution of Eq. 
(\ref{rg4}) is more complicated than the expression for spinless fermions 
given in Eq. (\ref{solvrg}) \cite{yue}. We will therefore not attempt to
make a quantitative comparison between our numerical results and the RG 
equations for the case of spin-1/2 electrons.

It is interesting to note that unlike in the spinless case, Eq. (\ref{rg4})
allows the possibility of $|t|$ increasing for a while and then decreasing (or
vice versa) \cite{yue}. This is due to the flow of the couplings $g_1$ and 
$g_2$; it happens if ${\tilde U} (0) - 2 {\tilde U} (2k_F)$ and ${\tilde U} 
(0) - (1/2) {\tilde U} (2k_F)$ have opposite signs. This is precisely the 
situation for the Hubbard model in which ${\tilde U} (0)$ and ${\tilde U} 
(2k_F)$ are both equal to the Hubbard parameter $U$. 

\section{Lattice models and the Hartree-Fock approximation}

Although the system of interest may be defined in the continuum, it is
convenient to approximate it by a lattice model in order to do numerical 
calculations. (We should of course ensure that physical quantities
like the wire length are much larger than the lattice spacing, so that
the lattice approximation does not introduce significant errors).
Let us discuss the form of the lattice Hamiltonians 
that we will consider. The Hamiltonian has a hopping term
\beq
H_0 ~=~ - \gamma ~\sum_{n,\sigma} ~[~ c_{n,\sigma}^\dagger c_{n+1,\sigma} ~+~
c_{n+1,\sigma}^\dagger c_{n,\sigma} ~] ~,
\label{hhop}
\eeq
where $c_{n,\sigma}$ annihilates an electron with spin $\sigma$ at site $n$,
and the hopping amplitude $\gamma$ will be taken to be positive for 
convenience. Next, we will place an impurity at one site, say, $n=0$, so that
\beq
H_V ~=~ V \sum_\sigma ~c_{0,\sigma}^\dagger c_{0,\sigma} ~.
\label{hv}
\eeq 
Finally, we will introduce an interaction between the electrons in
the wire (but not in the leads which will be considered in the next section).
For the spin-1/2 case, the interaction will be taken to be of the Hubbard form,
\beq
H_{\rm int} ~=~ U ~\sum_n ~c_{n,\ua}^\dagger c_{n,\ua} ~c_{n,\da}^\dagger 
c_{n,\da} ~.
\label{int1}
\eeq
For the case of spinless electrons, 
the spin index $\sigma$ will be dropped in Eqs. (\ref{hhop}) and (\ref{hv}),
and the interaction will be taken to be between nearest neighbor sites,
\beq
H_{\rm int} ~=~ \frac{U}{2} ~\sum_n ~c_n^\dagger c_n ~c_{n+1}^\dagger 
c_{n+1} ~.
\label{int2}
\eeq
(For spinless electrons, we cannot have an on-site interaction since
$(c_n^\dagger c_n)^2 = c_n^\dagger c_n$).

In the absence of the impurity and interactions, the energy is related to 
the wavenumber as
\beq
E(k) ~=~ -~ 2 \gamma ~\cos k ~,
\label{disp}
\eeq
where $- \pi < k \le \pi$. (We have set the lattice spacing equal to 1). The 
velocity is given by $v= dE/dk = 2 \gamma \sin k$. If the chemical potential 
is given by $\mu$, with $-2 \gamma < \mu < 2 \gamma$, the ground state is the 
one in which the states with momenta $-k_F$ to $k_F$ are filled, where $\mu 
= - 2 \gamma \cos k_F$, with $0 < k_F < \pi$. It is convenient to define the 
Fermi energy as the difference between the chemical potential and the bottom 
of the band, namely, $E_F = \mu + 2 \gamma = 2 \gamma (1 - \cos k_F)$.

Let us now consider the effect of the impurity placed at $n=0$ (we are still 
ignoring the interactions). The reflection and transmission amplitudes are 
given by
\bea
r(k) &=& - ~\frac{iV}{2 \gamma \sin k ~+~ iV} ~, \non \\
{\rm and} \quad t(k) &=& \frac{2 \gamma \sin k}{2 \gamma \sin k ~+~ iV} ~.
\label{rt2}
\eea
In addition, there is a bound state if $V < 0$; the bound state energy is
given by $E_b = - 2 \gamma \cosh k_b$, where $V = - 2 \gamma \sinh k_b$,
with $k_b > 0$. The normalized bound state wave function is given by
\beq
\psi_{b,n} ~=~ (\tanh k_b)^{1/2} ~e^{-k_b |n|} 
\label{psib}
\eeq
for all values of $n$.

It is interesting to compute the electron density 
$\rho_n \equiv \rho_{n,\ua} = \rho_{n,\da}$, as a function of the 
site index $n$. The contribution from the scattering states is given by
\bea
\rho_n &=& \int_0^{k_F} ~\frac{dk}{2\pi} ~[~ |e^{-ik|n|} ~+~ r(k) 
e^{ik|n|}|^2 ~ +~ |t(k)|^2 ~] \non \\
&=& {\bar \rho} ~+~ \int_0^{k_F} ~\frac{dk}{2\pi} ~[~ r(k) e^{i2k|n|} ~+~
r^*(k) e^{-i2k|n|} ~] ~, \non \\
& &
\eea
where ${\bar \rho} = k_F /\pi$ is the value of the density very far
from the impurity. In the limit $|n| \rightarrow \infty$, we find that
the density has oscillations given by
\bea
& & \rho_n ~-~ {\bar \rho} \non \\
& & \simeq~ -~ \frac{i}{4 \pi |n|} [~ r (k_F) e^{i2k_F|n|} ~-~ r^* (k_F) 
e^{-i2k_F|n|} ~] ~.
\label{rhon}
\eea

Now we will introduce interactions and consider the HF approximation for the 
cases of spin-1/2 and spinless electrons. This will give us another way of
understanding the RG equations discussed in Sec. II.

\subsection{Spin-1/2 electrons}

For the interaction given in Eq. (\ref{int1}), the HF approximation takes 
the form
\beq
H_{\rm int,HF} ~=~ U ~\sum_n ~[~ \rho_n ~(c_{n,\ua}^\dagger c_{n,\ua} ~+~ 
c_{n,\da}^\dagger c_{n,\da}) ~-~ \rho_n^2 ~]~,
\label{int3}
\eeq
where we have set $<c_{n,\ua}^\dagger c_{n,\ua}>= <c_{n,\da}^\dagger 
c_{n,\da}>= \rho_n$. [This is called the restricted HF. One can also make
an unrestricted HF approximation in which one allows $<c_{n,\ua}^\dagger 
c_{n,\ua}>$ to differ from $<c_{n,\da}^\dagger c_{n,\da}>$.]

At this point, we would like to impose the condition that the interaction 
should have no effect on the conductance if there is no impurity, i.e., 
if $\rho_n$ is equal to the constant $\bar \rho$ at all sites. The reason
for imposing this condition is that we do not want the interactions by
themselves to lead to any scattering; that would change the conductance
from the ideal value of $2e^2 /h$ in the absence of any impurities which is
an undesirable effect. (This will become clearer in the next section when
we describe our model for the wires and the leads). We will therefore 
modify Eq. (\ref{int3}) to the form
\bea
H_{\rm int,HF} &=& U ~\sum_n ~(\rho_n ~-~ {\bar \rho}) ~(c_{n,\ua}^\dagger 
c_{n,\ua} ~+~ c_{n,\da}^\dagger c_{n,\da}) ~, \non \\
& &
\label{int4}
\eea
where we have dropped the constant $\rho_n^2$ which turns out to have no 
effect on the conductance. (The terms proportional to $\bar \rho$ in Eq.
(\ref{int4}) are equivalent to adding a chemical potential).

To gain an insight into the effect of Eq. (\ref{int4}), let us consider the 
Born approximation. In order to use this approximation, we will assume here
that both $V$ and $U$ are small. The total potential seen by either spin-up or
spin-down electrons at site $n$ is
\beq
{\cal V}_n ~=~ V \delta_{n,0} ~+~ U (\rho_n ~-~ {\bar \rho}) ~.
\eeq
For small values of $|V|/v_F$ (where $v_F = 2 \gamma \sin k_F$ is the 
Fermi velocity), $r(k_F) = -iV/v_F$ from Eq. (\ref{rt2}), and we have 
\beq
{\cal V}_n ~=~ V \delta_{n,0} ~-~ \frac{VU}{4 \pi v_F |n|} ~ (e^{i2k_F|n|} ~
+~ e^{-i2k_F|n|}) ~,
\eeq
where the second term is valid only for $|n| >> \pi /k_F$.

The Born approximation for the reflection amplitude in one dimension for
a lattice model is given by
\beq
r_B (k_F) ~=~ - \frac{i}{v_F} ~\sum_n {\cal V}_n ~e^{i2k_F n} ~.
\eeq
If the region in which the electrons interact with each other only
extends from $n=-L/2$ to $L/2$, then for $L>> \pi /k_F$, we get 
\beq
r_B (k_F) ~=~ - ~\frac{iV}{v_F} ~[~ 1 ~-~ \frac{U}{2 \pi v_F} ~\ln 
\frac{L}{2} ~]~.
\label{rb1}
\eeq
If $U>0$, we see that the conductance $G = (2e^2 /h) [1-|r_B (k_F)|^2]$ 
increases with $L$ till it reaches a maximum when $\ln (L/2) \simeq 2 \pi 
v_F/U$. However, the Born approximation that we are using here cannot be 
trusted up to such large length scales because we are not self-consistently 
modifying the densities at the different sites in response to the 
renormalization of the reflection amplitudes (produced by the interaction 
$U$). Because of this lack of self-consistency, we can only trust the Born 
approximation results up to length scales where the reflection amplitude 
has not changed very much from the value of $-iV/v_F$ that it has in the 
non-interacting theory. 

We note that Eq. (\ref{rb1}) is consistent with Eq. (\ref{rg4}) for
small values of $V$ and $(U/2\pi v_F) \ln l$, since $(g_2 (l)-2g_1 (l))/2\pi 
v_F = - U/2\pi v_F$ for such values of $l$.

\subsection{Spinless electrons}

For the interaction given in Eq. (\ref{int2}), the HF approximation takes 
the form
\bea
H_{\rm int,HF} ~=~ \frac{U}{2} ~\sum_n ~[~ & & \rho_{n+1} ~c_n^\dagger c_n ~
+~ \rho_n c_{n+1}^\dagger c_{n+1} \non \\
& & -~ <c_n^\dagger c_{n+1}> c_{n+1}^\dagger c_n \non \\
& & -~ <c_{n+1}^\dagger c_n> c_n^\dagger c_{n+1} ~]~,
\label{int5}
\eea
where $\rho_n \equiv <c_n^\dagger c_n>$, and we have dropped some constants.
In the absence of any impurities, $\rho_n = k_F /\pi$ and $<c_{n+1}^\dagger 
c_n> = (\sin k_F)/\pi$ for all values of $n$.

Once again, we impose the condition that the interaction should have no 
effect on the conductance if there is no impurity. We therefore 
modify Eq. (\ref{int5}) to the form
\bea
H_{\rm int,HF} ~=~ \frac{U}{2} ~\sum_n ~[ & & (\rho_{n+1} ~+~ \rho_{n-1} ~-~ 
2 {\bar \rho}) ~c_n^\dagger c_n \non \\
& & -~ (<c_n^\dagger c_{n+1}> - \frac{\sin k_F}{\pi}) ~c_{n+1}^\dagger c_n 
\non \\
& & -~ (<c_{n+1}^\dagger c_n> - \frac{\sin k_F}{\pi}) ~c_n^\dagger c_{n+1} ~]~,
\non \\
& & 
\label{int6}
\eea
where ${\bar \rho} = k_F /\pi$ as before.

Let us again use the Born approximation to understand the effect of Eq. 
(\ref{int6}), assuming that both $V$ and $U$ are small. If one adds
a perturbation of the form
\bea
H_{\rm pert} ~=~ \sum_n ~[~ & & {\cal V}_n ~c_n^\dagger c_n \non \\
& & +~ {\cal V}_{n+1/2} ~c_{n+1}^\dagger c_n ~+~ {\cal V}_{n+1/2}^* ~
c_n^\dagger c_{n+1} ~] \non \\
& & 
\eea
to the Hamiltonian in Eq. (\ref{hhop}), the Born approximation for the
reflection amplitude is given by
\bea
r_B (k_F) = - \frac{i}{v_F} \sum_n ~[ & & {\cal V}_n ~e^{i2k_F n} \non \\
& & + ({\cal V}_{n+1/2} + {\cal V}_{n+1/2}^*) ~e^{ik_F (2n+1)} ]~. \non \\
& &
\label{rb2}
\eea
In our case,
\bea
{\cal V}_n &=& V \delta_{n,0} ~+~ \frac{U}{2} ~[ ~\rho_{n+1} ~+~ \rho_{n-1} ~
-~ 2 {\bar \rho} ~] ~, \non \\
{\cal V}_{n+1/2} &=& -~ \frac{U}{2} ~[~ <c_n^\dagger c_{n+1}> ~-~ 
\frac{\sin k_F}{\pi} ~]~.
\eea
Hence,
\bea
{\cal V}_n ~=~ V \delta_{n,0} ~-~ \frac{iU \cos(2k_F)}{4 \pi |n|} ~ & &
[~ r (k_F) e^{i2k_F|n|} \non \\ 
& & ~-~ r^* (k_F) e^{-i2k_F|n|} ~] ~, \non \\
& &
\eea
where the second term is only valid for $|n| \rightarrow \infty$, and
\bea
{\cal V}_{n+1/2} &=& \frac{iU}{8\pi |n|} ~[~ r (k_F) ~e^{i2k_F |n| + 
i {\rm sgn} (n) k_F} \non \\
& & ~~~~~~~~~-~ r^* (k_F) ~e^{-i2k_F |n| - i {\rm sgn} (n) k_F} ~] ~, 
\non \\
& &
\eea
for $|n| \rightarrow \infty$; here ${\rm sgn} (n) \equiv n /|n|$. 
For small values of $|V|/v_F$, $r (k_F) = 
- i V /v_F$. Substituting everything in the Born approximation (\ref{rb2}), 
and assuming that the region in which the electrons interact with each other 
extends from $n=-L/2$ to $L/2$, where $L>> \pi /k_F$, we find that
\beq
r_B (k_F) ~=~ - ~\frac{iV}{v_F} ~[~ 1 ~+~ \frac{U}{2 \pi v_F} ~(1 - \cos 
(2k_F)) ~\ln \frac{L}{2} ~]~.
\label{rb3}
\eeq
Eq. (\ref{rb3}) is consistent with Eq. (\ref{rg3}) for small values of $V$ 
and $(U/2\pi v_F) \ln l$, since $\alpha = (U/ 2\pi v_F)[1 - \cos (2k_F)]$.

\section{Non-equilibrium Green's function formalism}

In this section, we will introduce the NEGF formalism which will allow us to 
study the conductance of a wire with any length, both short (of the order 
of $\pi /k_F$) and long (where a continuum description and RG analysis may be 
expected to be reliable). 

\subsection{Self-energy, density and conductance}

An important concept in the NEGF formalism is a ``self-energy" which 
describes the amplitude for an electron to leave or enter the wire from the 
leads (reservoirs) which are maintained at some 
chemical potentials and temperature \cite{datta,meir,datta2,dhar,tsukada}.
The self-energy is a non-Hermitian term in the single-particle Hamiltonian of 
the wire. To see how this arises, let us begin by modeling
one of the reservoirs by a tight-binding Hamiltonian
\beq 
H ~=~ -\gamma ~\sum_{n,\sigma} ~(~c_{n,\sigma}^\dagger c_{n+1,\sigma} ~ 
+~ c_{n+1,\sigma}^\dagger c_{n,\sigma} ~) ~. 
\eeq
The energy of an electron in the reservoir is related to its wavenumber as 
$E = - 2 \gamma \cos k$. The reservoir has a chemical potential $\mu$ and 
an inverse temperature $\beta = 1/(k_B T)$. The reservoir is semi-infinite;
the last site at one end of the reservoir is coupled to the first site of
the wire by a hopping amplitude $\gamma'$ (see Fig. 1).

\begin{figure}[htb]
\begin{center}
\epsfig{figure=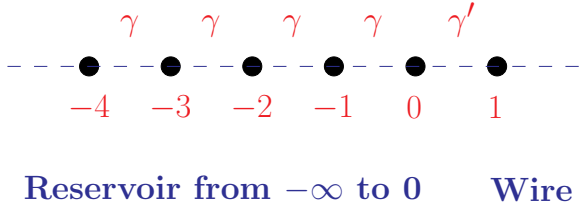,width=8.5cm}
\end{center}
\caption{(Color online) Picture of a semi-infinite reservoir going from site
number 0 to $- \infty$ and a wire beginning from site number 1.}
\end{figure}

The reservoir gives rise to a self-energy at the first site of the wire of 
the form \cite{datta,dhar}
\beq 
\Sigma (E) ~=~ \sigma (E) ~\sum_\sigma ~c_{1,\sigma}^\dagger c_{1,\sigma} ~,
\label{self}
\eeq
where
\beq 
\sigma (E) ~=~ -~ \frac{\gamma'^2}{\gamma} ~e^{ik} 
\eeq
for $-2\gamma \le E = - 2 \gamma \cos k \le 2\gamma$ (with $0 \le k \le \pi$),
\beq 
\sigma (E) ~=~ -~ \frac{\gamma'^2}{\gamma} ~e^{-k} ~-~ i \eta 
\eeq
for $E = - 2 \gamma \cosh k \le -2\gamma$ (with $k \ge 0$), and
\beq 
\sigma (E) ~=~ \frac{\gamma'^2}{\gamma} ~e^{-k} ~-~ i \eta
\eeq
for $E = 2 \gamma \cosh k \ge 2\gamma$ (with $k \ge 0$). Here $\eta$ is
an infinitesimal positive number which appears only if $E$ lies 
outside the range $[-2\gamma, 2\gamma]$. For $-2 \gamma < E < 2 \gamma$,
the self-energy already has an imaginary piece, so it is not necessary to
add an $i \eta$ term.

\begin{figure}[htb]
\epsfig{figure=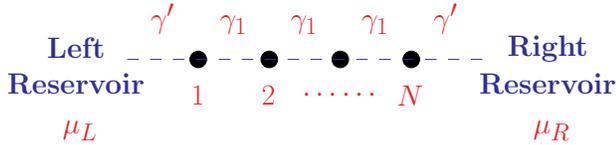,width=8.5cm}
\caption{(Color online) A quantum wire system with a $N$-site wire in the 
middle and semi-infinite reservoirs on its left and right.}
\end{figure}

Now let us consider the complete system which consists of a wire with $N$ sites
in the middle, and reservoirs on its left and right as shown in Fig. 2. If the
wire is also modeled by a tight-binding Hamiltonian (with a hopping amplitude
$\gamma_1$), the Green's function of the wire at energy $E$ is given by a 
$N \times N$ matrix
\bea 
& & G (E) ~=~ [~ E ~I ~-~ H ~-~ \Sigma_L (E) ~-~ \Sigma_R (E) ~]^{-1} ~, 
\non \\ 
& & \non \\
& & H ~+~ \Sigma_L (E) ~+~ \Sigma_R (E) ~= \non \\
& & \non \\
& & -~ \left( \begin{array}{ccccc} ~\\
\sigma (E)~ & ~ \gamma_1~ & ~0~ & ~\cdots~ & ~\cdots~ \\
~\gamma_1~ & ~0~ & ~\gamma_1~ & ~0~ & ~\cdots~ \\
~0~ & ~\gamma_1~ & ~0~ & ~\gamma_1~ & ~0~ \\
~\cdots~ & ~\cdots~ & ~\cdots~ & ~\cdots~ & ~\cdots~ \\
~\cdots~ & ~\cdots~ & ~0~ & ~\gamma_1~ & ~\sigma (E)~ \end{array} \right) ~. 
\eea
(The effects of the reservoirs on the wire is completely taken into account
by the self-energy terms). We would like to emphasize that the relation 
between $E$ and $k$ is given entirely by the dispersion in the reservoirs 
(we assume that the dispersion is the same in both the reservoirs), and 
{\it not} on the form of the Hamiltonian $H$ inside the wire.

The density matrices due to electrons coming in from 
the left and right reservoirs are given by \cite{datta,meir,datta2,dhar}
\beq 
\int ~\frac{dE}{2\pi} ~G \Gamma_L G^\dagger ~f_L \quad {\rm and} \quad 
\int ~\frac{dE}{2\pi} ~G \Gamma_R G^\dagger ~f_R 
\label{density}
\eeq
respectively, where $\Gamma_a (E) = i (\Sigma_a - \Sigma_a^\dagger),$
\beq f_a (E) ~=~ [e^{\beta (E - \mu_a)} + 1]^{-1} \eeq
is the Fermi function, and $\mu_a~$ is the chemical potential in reservoir $a$.
We note that if $E$ lies outside the range $[-2\gamma,2\gamma]$, the matrix
$\Gamma_a (E) \sim 2 \eta$ is infinitesimal. In that case, the density matrix 
can still receive a contribution from certain states; these are typically bound
states which have a discrete set of energies $E_b$. The reason that such states
can make a finite contribution in Eq. (\ref{density}) even though $\Gamma_a$
is infinitesimal is that $G$ takes the form $1/(E - E_b +i \eta)$, and
\beq
{\rm Lim}_{\eta \rightarrow 0^+} ~\frac{2~\eta}{(E-E_b)^2 ~+~ 
\eta^2} ~=~ 2~\pi~\delta (E-E_b) ~.
\eeq 

Finally, the current is given by the expression
\beq 
I ~=~ - ~\frac{e}{h} ~\int ~dE ~{\rm trace} (G \Gamma_L G^\dagger \Gamma_R) ~
[ f_L (E) - f_R (E) ] ~, 
\label{current}
\eeq
and the conductance is $G ~=~ e ~I / (\mu_L ~-~ \mu_R) ~.$ If $E$ lies outside
the range $[-2\gamma,2\gamma]$, $\Gamma_L (E)$ and $\Gamma_R (E)$ are both 
infinitesimal, and the current does not get any contribution from states lying
in that energy range. (The difference between the density matrix and the
current for such states is that the density matrix has only one factor of 
$\eta$ in the numerator while the current has two factors of $\eta$).

We will be interested below in the case of linear response, i.e, the limit
$\mu_L \rightarrow \mu_R = \mu$. In that limit, the Fermi wavenumber $k_F$ is
given by $\mu = - 2 \gamma \cos k_F$. Further, the conductance takes the form
\bea
G &=& \frac{e^2}{h} ~\int_{-2\gamma}^{2\gamma} ~dE ~{\rm trace} (G \Gamma_L 
G^\dagger \Gamma_R) ~\frac{\beta}{[2 ~\cosh \frac{\beta}{2} (E - \mu)]^2}. 
\non \\
&&
\eea
(We have to multiply the above expression by 2 for spin-half electrons).

\subsection{Interactions}

Let us now consider how interactions can be studied within the NEGF formalism.
Note that this formalism works with a one-particle Hamiltonian, e.g., the 
self-energy in Eq. (\ref{self}) is given in terms of the energy of a single 
electron which is entering or leaving the wire. Hence, the only way to deal 
with interactions is to do a HF decomposition as shown in Eqs. (\ref{int3}) 
and (\ref{int5}); these take the form of corrections to the on-site chemical 
potential or the hopping amplitude.

However, there is a difficulty in using the expressions in 
(\ref{int3}) or (\ref{int5}). We want to have 
interactions between the electrons only in the wire, not in the reservoirs.
The HF decompositions shown in (\ref{int3}) and (\ref{int5}) will then lead to
one-body terms only in the wire; these terms will back-scatter electrons coming
from the reservoirs, and hence reduce the conductance from its
ideal value of $2e^2 /h$ or $e^2 /h$. We would like to ensure that there is 
no such scattering if there are no impurities inside the wire and if
the leads connect adiabatically to the wires, i.e., if the entire system with
wire and reservoirs is translation invariant. We know that 
two-body scatterings between electrons conserve momentum and therefore do not
affect the conductance in the absence of impurities and scattering from the
lead-wire junctions. We therefore modify the form of the HF decomposition from
Eqs. (\ref{int3}) and (\ref{int5}) to Eqs. (\ref{int4}) and (\ref{int6}) 
respectively; the extra terms in those equations ensure that the HF terms 
vanish identically if there are no impurities. [An alternative way of ensuring
adiabaticity between the leads and the wires is to turn on the interaction 
strength $U$ smoothly from zero in the leads to a finite value inside 
the wire. This, however, is harder to implement numerically. Subtracting the 
mean values as in (\ref{int4}) and (\ref{int6}) is easier to implement, and
it serves the same purpose because the subtracted quantities approach zero 
near the ends of the wires (this will become clear from the numerical 
results presented below).]

In the presence of interactions, we will implement a self-consistent NEGF 
calculation as follows:
\begin{itemize} 
\item{Start with the Hamiltonian with no interactions, and calculate the 
density matrix. The diagonal elements of the density matrix give the 
densities at different sites.} 
\item{Use the HF approximation to compute the Hamiltonian with interactions, 
and use that to calculate the density matrix again.} 
\item{Repeat the previous step till the density changes no further,
i.e., between two successive iterations, the maximum change in density on 
any lattice site is less than about $10^{-4}$.} 
\item{Use the converged density to compute the conductance.} 
\end{itemize}

\section{Numerical results}

In this section, we will present the results obtained by the NEGF formalism, 
first for non-interacting electrons, and then for interacting spin-1/2 and 
spinless electrons, using the procedure described in Sec. IV B. We will study 
the dependence of the conductance $G$ on the wire length $L$, the inverse 
temperature $\beta$, the impurity strength $V$, and the interaction parameter 
$U$. Finally, we will make a quantitative comparison between our numerical
results and the RG equations for the case of spinless electrons.

We present some details about our choice of parameters for the calculations.
We always take the wire to have an odd number of sites, and the impurity to be
on the middle site. The hopping amplitudes in the reservoirs and in the wire 
is chosen to be the same ($\gamma = \gamma_1 = \gamma' =1$); the values of
the inverse temperature $\beta = 1/(k_B T)$ is quoted in units of $1/\gamma$.
(We set both $\hbar$ and the lattice spacing equal to 1). We choose $k_F = 
\pi /10$; hence, $v_F = 2 \sin k_F = 0.618$, $\mu = - 2 \cos k_F = -1.902$, 
and $E_F = 2 (1 - \cos k_F) = 0.098$. The thermal coherence length is
given by $L_T = 0.618 \beta$.
In the energy integrations for the current and conductance, we take the energy
step size to be $dE=2 \times 10^{-4}$. In the integration outside the energy 
range $[-2\gamma , 2 \gamma]$, we take the quantity $\eta$ to be 5 times $dE$.
Finally, in all the figures, the conductance $G$ is expressed in units of 
$2e^2 /h$ and $e^2 /h$ for the spin-1/2 and spinless cases respectively.

Our choice of parameters was dictated partly by experiments on quantum wires 
in semiconductor heterojunctions \cite{tarucha,liang,bkane,yacoby,aus,reilly},
and partly by our numerical limitations (which prevent us from going to
very large wire lengths). Experimentally, the ratio of the wire length to
the de Broglie wavelength of the electrons ($Lk_F /\pi$) ranges from 
about 20 to 200, the ratio of the wire length to the inverse temperature 
$L/\beta$ ranges from about 1/2 to 10, and the parameter $U/(2 \pi v_F)$ 
appearing in the Eq. (\ref{rb1}) is about $0.2$ - $0.3$ \cite{lal1}. In our 
calculations, $Lk_F /\pi$ goes from about 1 to 30, $L/\beta$ goes from about 
1/40 to 6, and the interaction parameters take the values $U/(2\pi v_F) = 
0.0773$ in the spin-1/2 case (Eq. (\ref{rb1})) and $(U/ 2 \pi v_F)[1 - 
\cos(2k_F)] = 0.0148$ in the spinless case (Eq. (\ref{rb3})) for $U=0.3$.
The much smaller value of the interaction parameter in the spinless case 
leads to smaller changes in the conductance in that case as we will see.

\subsection{Non-interacting electrons}

Let us first discuss some properties of a system in which there is a 
$\delta$-function impurity of strength $V$ in the middle of the wire, and 
there are no interactions between the electrons. Fig. 3 shows the Friedel 
oscillations in the density $<c_n^\dagger c_n>$ for $V=0.3$ for two different 
temperatures. Note that the density at the middle site is much lower than the 
mean value of ${\bar \rho} = k_F /\pi = 1/10$ due to the repulsive nature of 
the impurity. We also observe that the oscillations die out beyond a length 
scale of order $L_T = 0.618 \beta$ in the lower figure. This is a simple 
illustration of the idea of a thermal coherence length. 

\begin{figure}[htb]
\begin{center}
\epsfig{figure=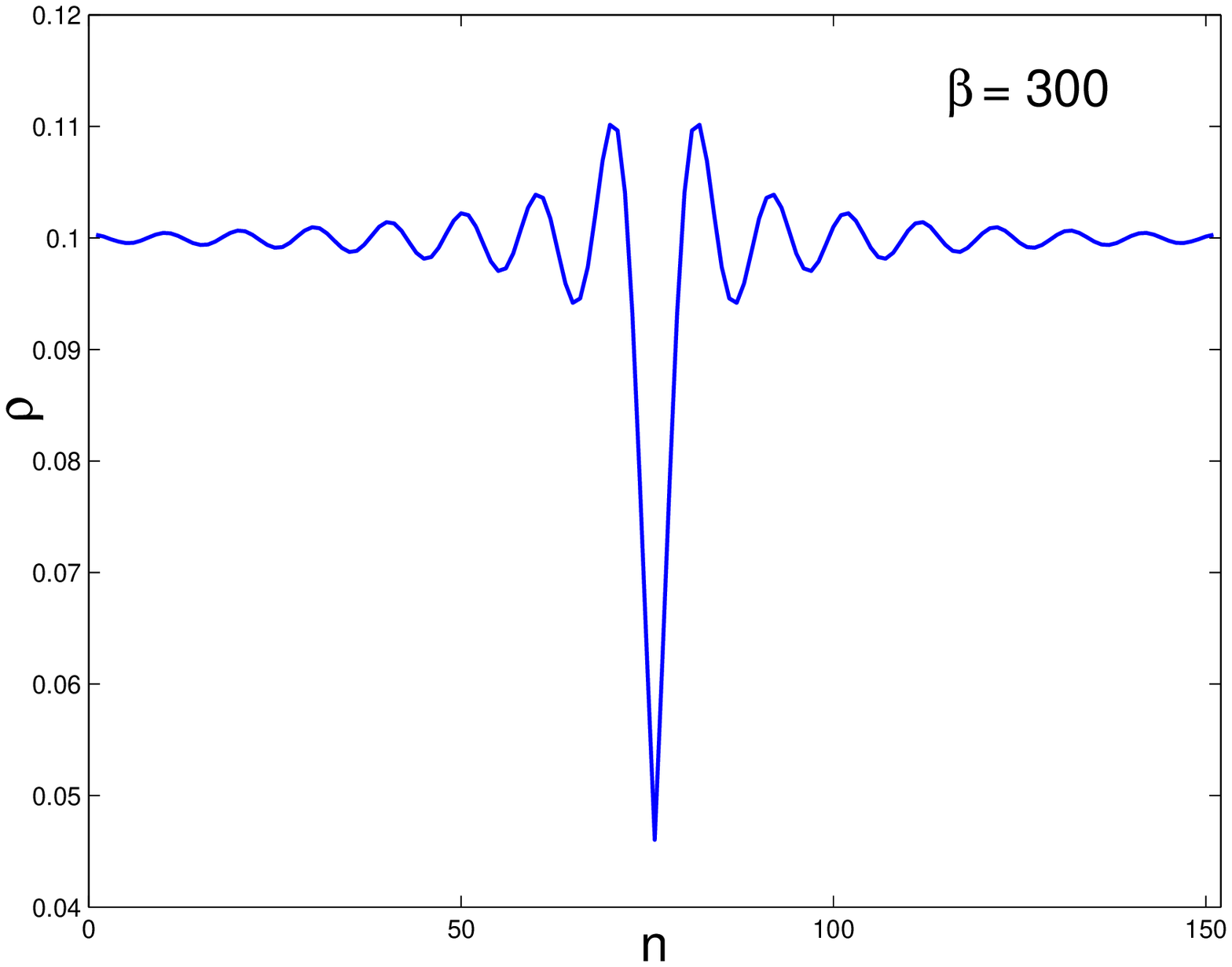,width=8.5cm}
\epsfig{figure=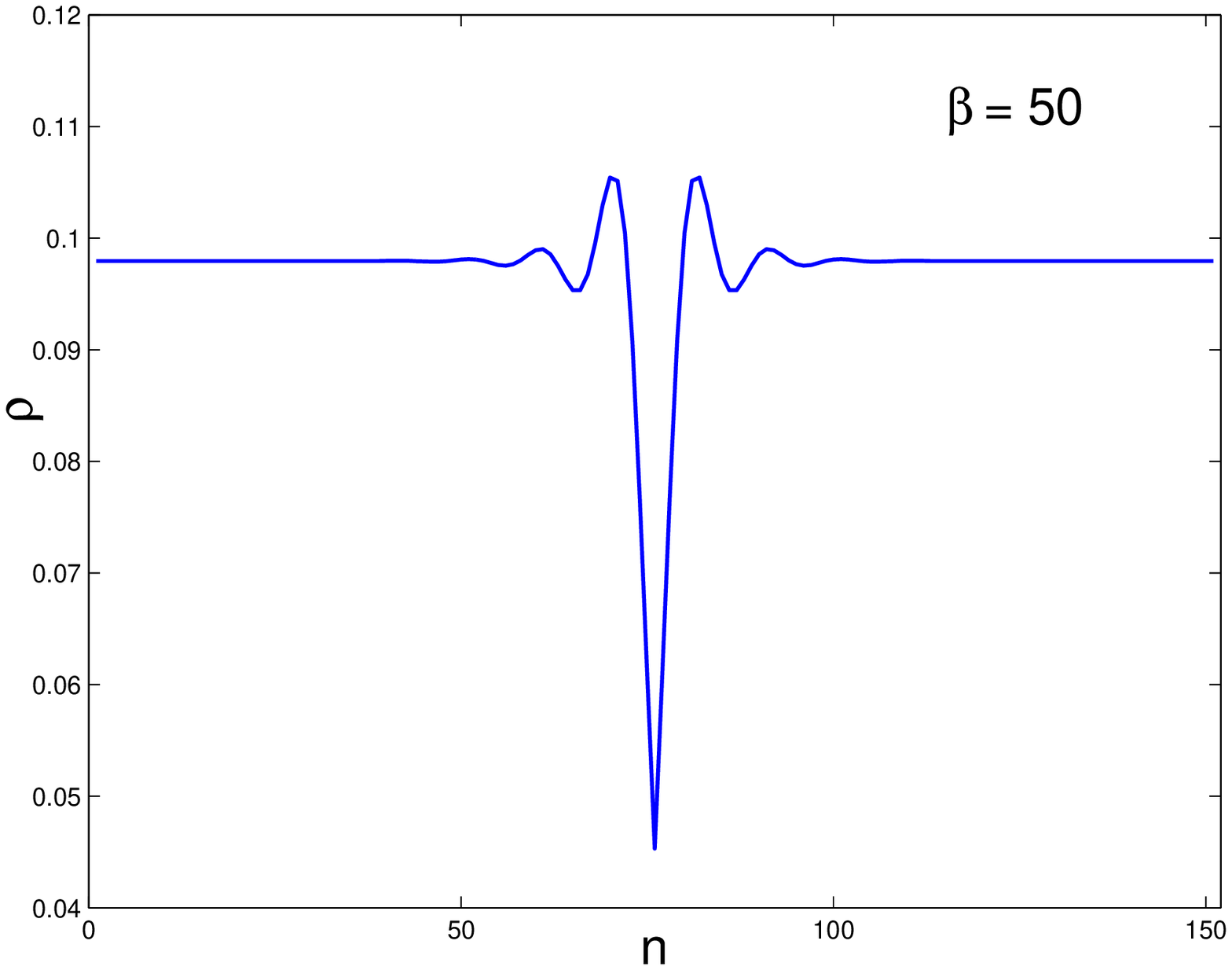,width=8.5cm}
\end{center}
\caption{(Color online) Friedel oscillations in the density of 
non-interacting electrons caused by an impurity with $V=0.3$ placed in the 
middle of a wire with 151 sites. The upper and lower figures have $\beta =$ 
300 and 50 respectively.}
\end{figure}

\begin{figure}[htb]
\begin{center}
\epsfig{figure=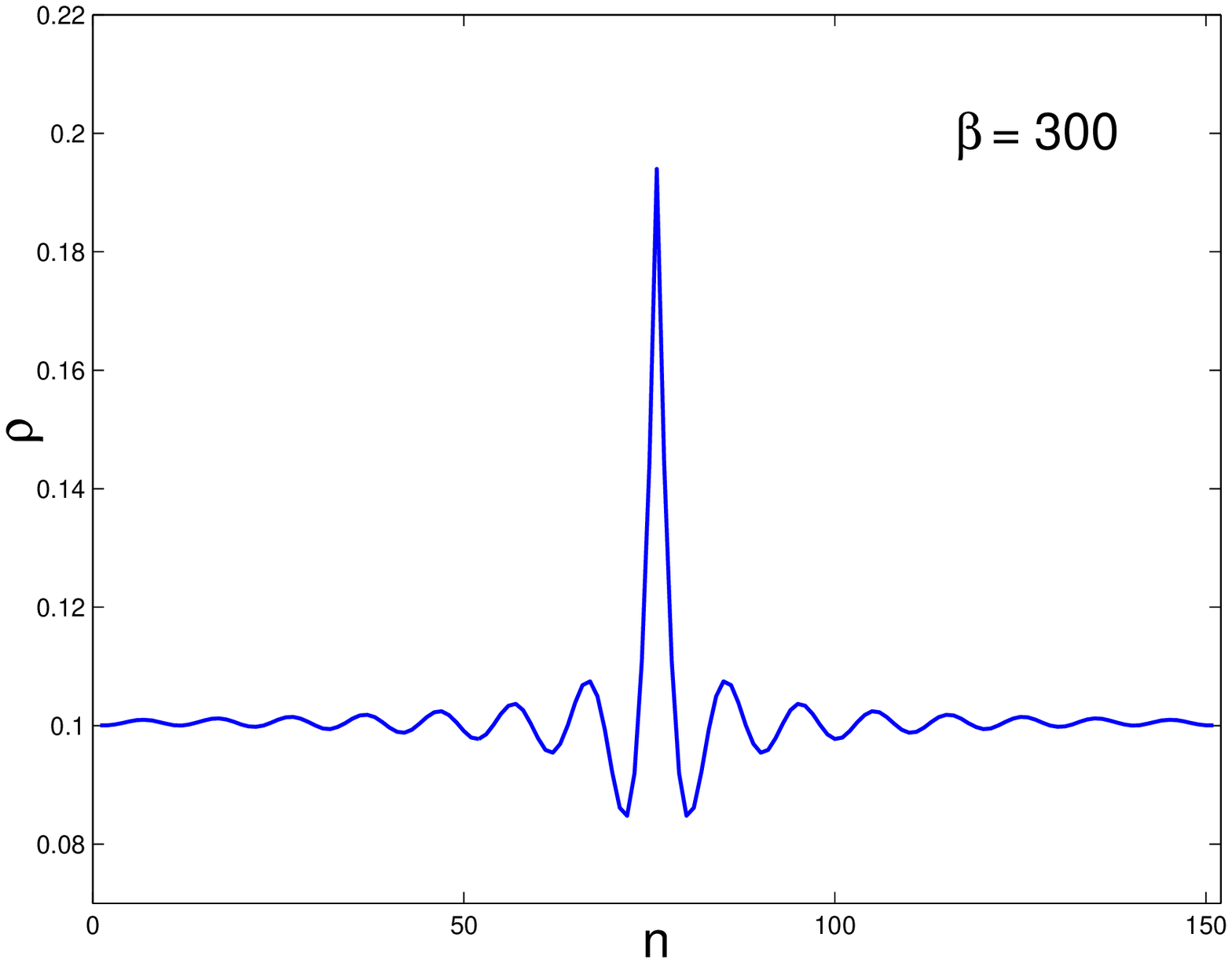,width=8.5cm}
\end{center}
\caption{(Color online) Friedel oscillations in the density of 
non-interacting electrons caused by an impurity with $V=-0.3$ placed in 
the middle of a wire with 151 sites and $\beta = 300$.} 
\end{figure}

\begin{figure}[htb]
\begin{center}
\epsfig{figure=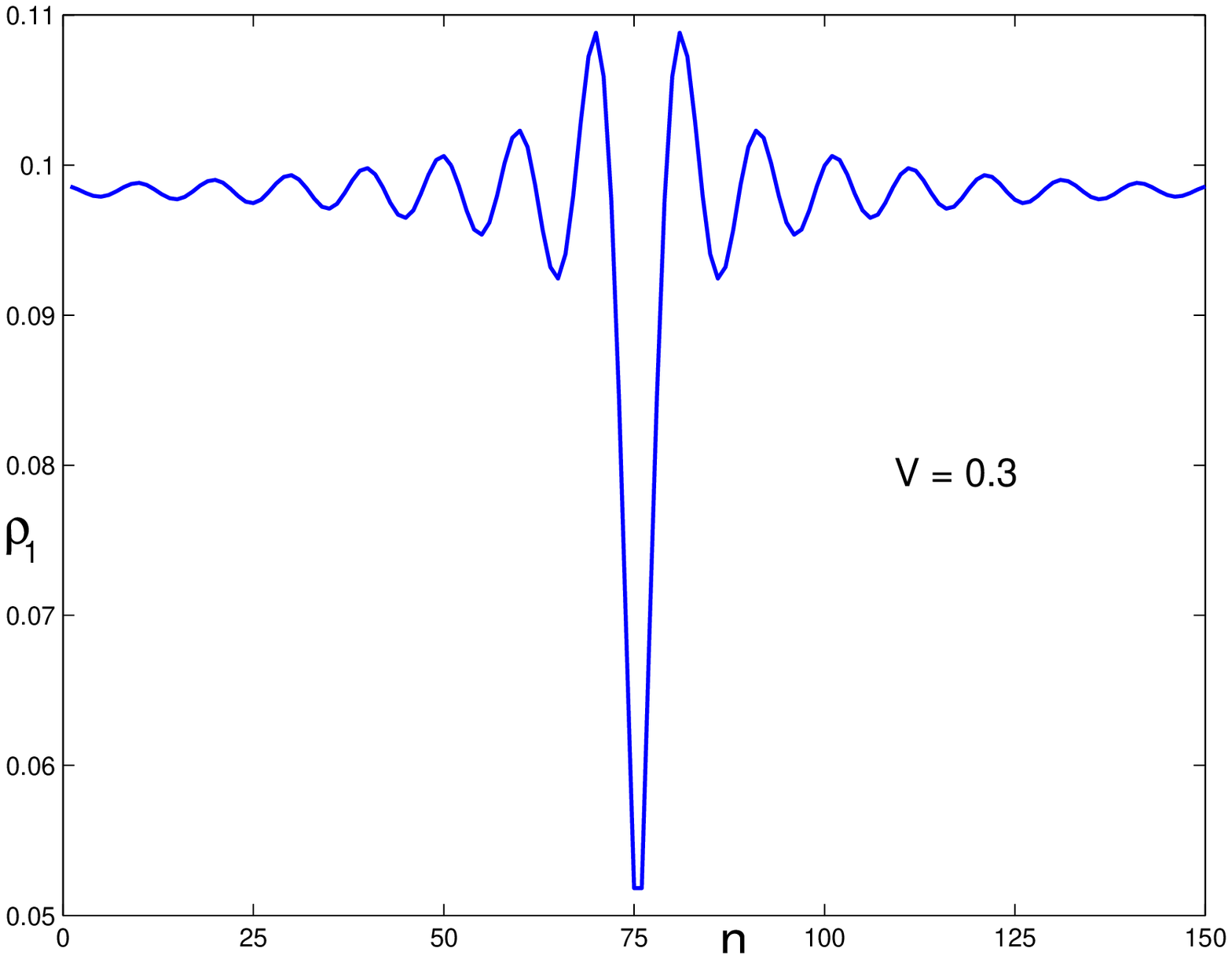,width=8.5cm}
\epsfig{figure=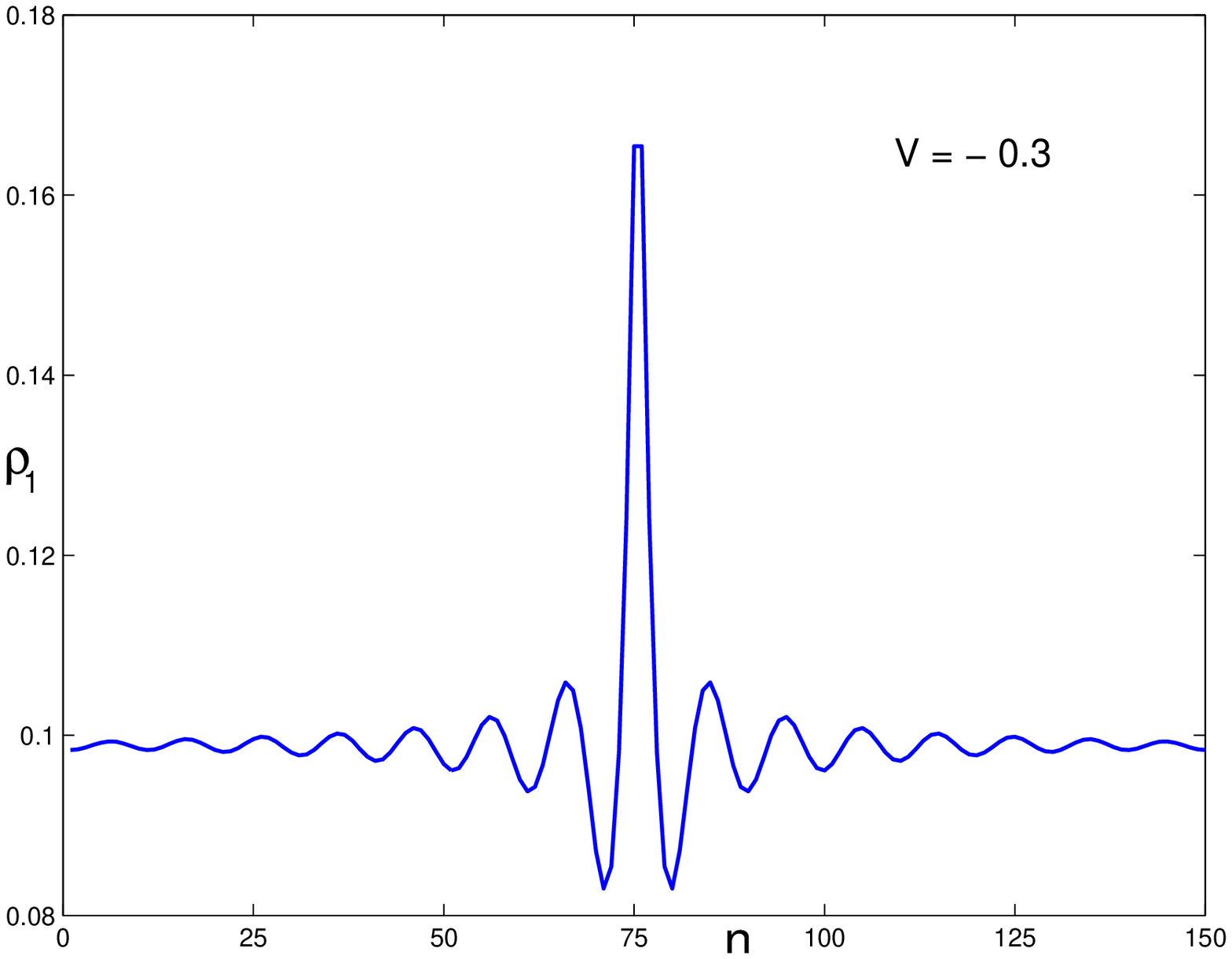,width=8.5cm}
\end{center}
\caption{(Color online) Plots of $\rho_1 \equiv <c_n^\dagger c_{n+1}>$ 
for non-interacting electrons in the presence of an impurity with $V=0.3$ 
(upper figure) and $-0.3$ (lower figure) placed in the middle of a wire with 
151 sites, with $\beta = 300$.} 
\end{figure}

\begin{figure}[htb]
\begin{center}
\epsfig{figure=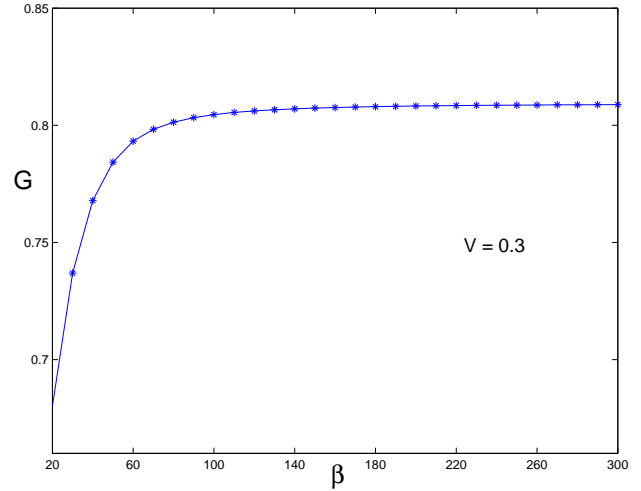,width=8.6cm}
\end{center}
\caption{(Color online) Conductance versus $\beta$ for non-interacting 
electrons in the presence of an impurity with $V=0.3$ placed in the middle 
of a wire with 201 sites.} 
\end{figure}

\begin{figure}[htb] 
\begin{center}
\epsfig{figure=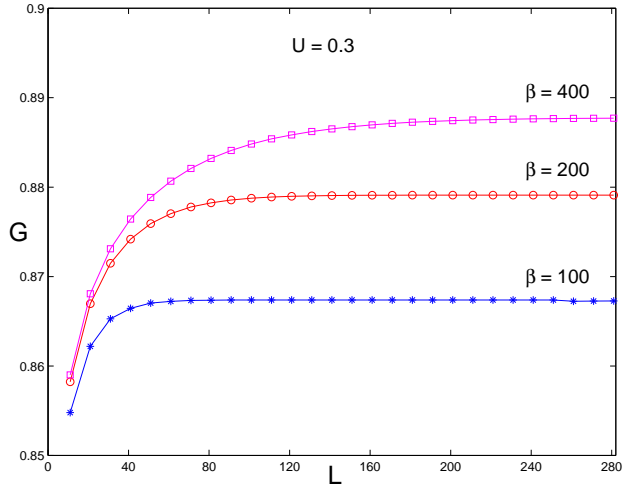,width=8.5cm}
\epsfig{figure=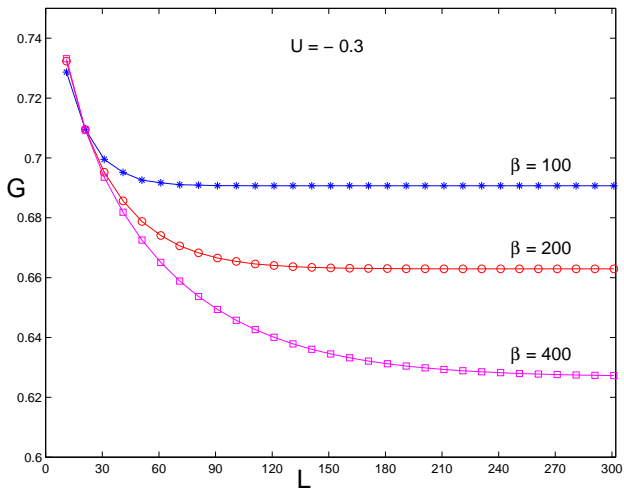,width=8.5cm}
\end{center}
\caption{(Color online) Conductance versus wire length for interacting 
spin-1/2 electrons for three different values of $\beta$, for $V=0.3$. $U=$ 
0.3 and -0.3 in the upper and lower figures respectively.}
\end{figure}

\begin{figure}[htb]
\begin{center}
\epsfig{figure=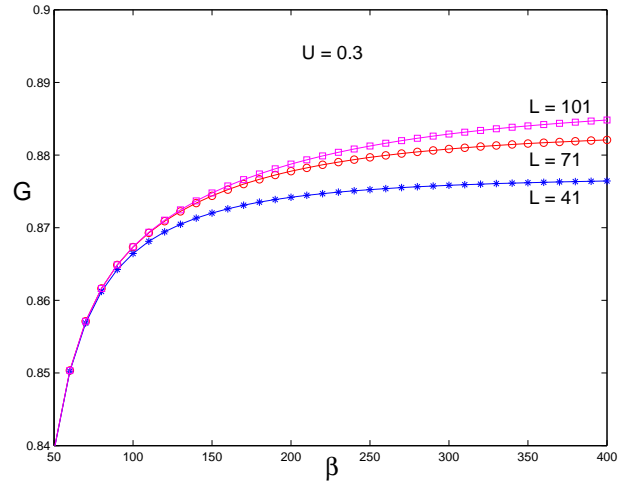,width=8.5cm}
\end{center}
\caption{(Color online) Conductance versus $\beta$ for interacting spin-1/2 
electrons for three different wire lengths, for $V=0.3$ and $U=0.3$.}
\end{figure}

\begin{figure}[htb]
\begin{center}
\epsfig{figure=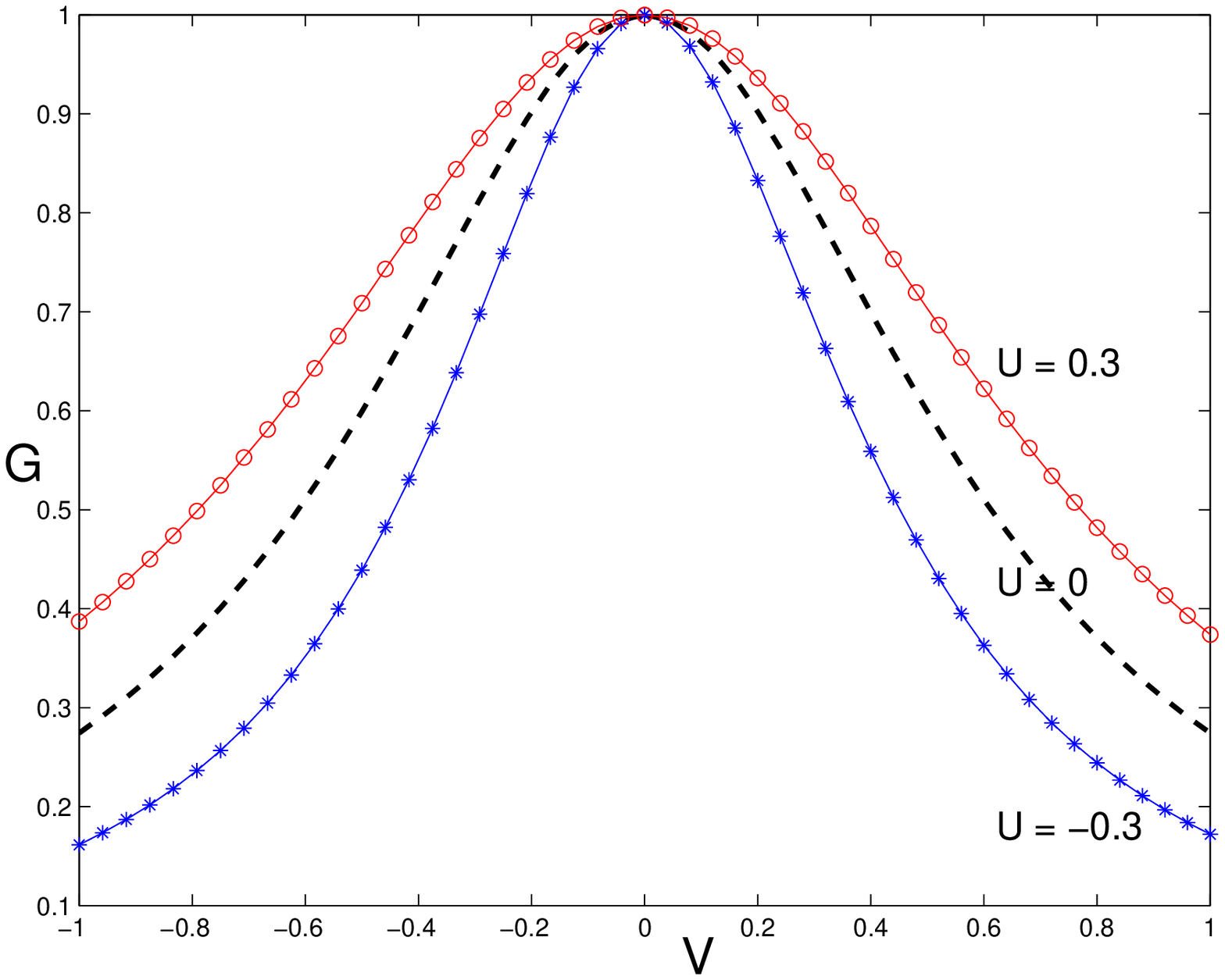,width=8.5cm}
\end{center}
\caption{(Color online) Conductance $G$ versus $V$ for interacting spin-1/2 
electrons for $U$ = 0.3, 0 and -0.3, with $L=101$ and $\beta = 100$.}
\end{figure}

\begin{figure}[htb]
\begin{center}
\epsfig{figure=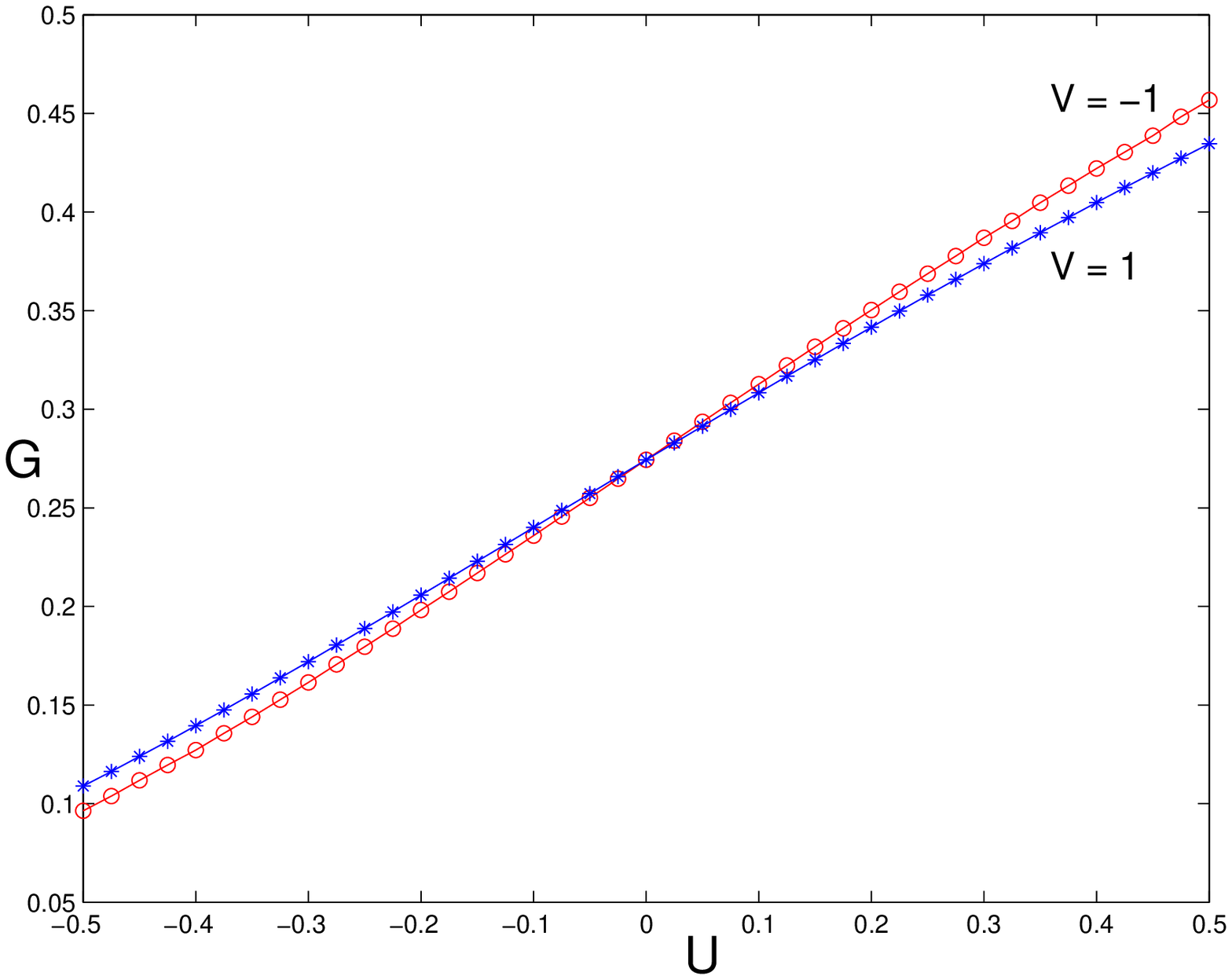,width=8.5cm}
\end{center}
\caption{(Color online) Conductance $G$ versus $U$ for interacting spin-1/2 
electrons for $V$ = 1 and -1, with $L=101$ and $\beta = 100$.}
\end{figure}

Fig. 4 shows the Friedel oscillations in the density for $V=-0.3$ for
$\beta = 300$. The density in the middle site is now much higher than 
the mean value; this is because of the presence of a bound state whose wave 
function has a peak at that site. Indeed, in the absence of interactions,
the difference in the site densities for $V=0.3$ and $-0.3$ is given entirely 
by the bound state. We see from Figs. 3 and 4 that the density difference 
at the site of the impurity is about 0.15; this agrees well with the
value of $\tanh k_b = 0.148$ given by Eq. (\ref{psib}). (For $V=-0.3$, we
have $k_b = 0.149$).

Finally, Fig. 5 shows the quantity $\rho_1 \equiv <c_n^\dagger c_{n+1}>$ for
$V=0.3$ and $-0.3$ for $\beta = 300$. There is a strong similarity between 
these and the density plots shown in Figs. 3 (upper figure) and 4. This
is because the electrons at the Fermi energy (which dominate the long 
distance properties of the system) have a wavelength which is $\pi /k_F = 10$
times longer than the lattice spacing; hence there is very little difference 
between $<c_n^\dagger c_n>$ and $<c_n^\dagger c_{n+1}>$.

For non-interacting electrons, the conductance depends on the temperature, 
but not on the length of the wire. For a $\delta$-function impurity of 
strength $V$, the conductance for spinless electrons is given by
\bea
G &=& \frac{e^2}{h} ~\int_{-2}^2 ~dE ~{\cal T} (E)~
\frac{\beta}{[2 ~\cosh \frac{\beta}{2} (E - \mu)]^2} ~, \non \\
{\cal T} (E) &=& \frac{4 \sin^2 k}{4 \sin^2 k + V^2}~.
\label{cond}
\eea
where $E= - 2 \cos k$. For $k_B T << E_F = \mu + 2$, Eq. (\ref{cond}) has 
a Sommerfeld expansion of the form \cite{ashcroft}
\beq
G ~=~ \frac{e^2}{h} ~[~ T(\mu) ~+~ \frac{\pi^2}{6} ~(k_B T)^2 ~T''(\mu) ~+~ 
\cdots ~]~.
\label{g0}
\eeq
For $k_F = \pi /10$ and $V=0.3$, $T(\mu) = 0.809$ and $(\pi^2 /6) T''(\mu) = 
42.1$; this implies a significant temperature dependence of the conductance. 
Fig. 6 shows that the conductance changes appreciably with $\beta$ till 
$\beta$ reaches about 100. We have to keep this in mind when studying the 
temperature dependence of the conductance of a system of interacting electrons.

\begin{figure}[h!]
\begin{center}
\epsfig{figure=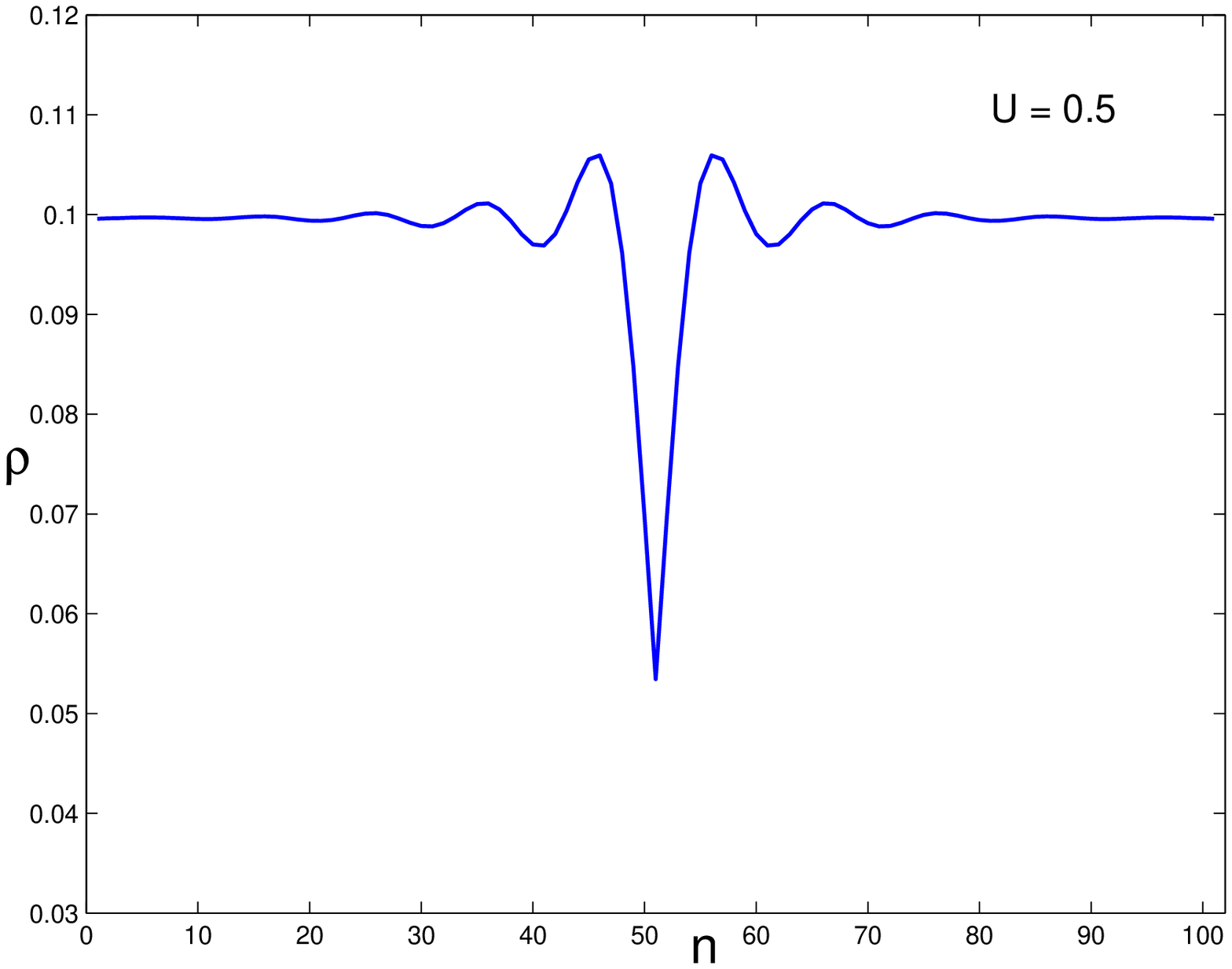,width=8.5cm}
\epsfig{figure=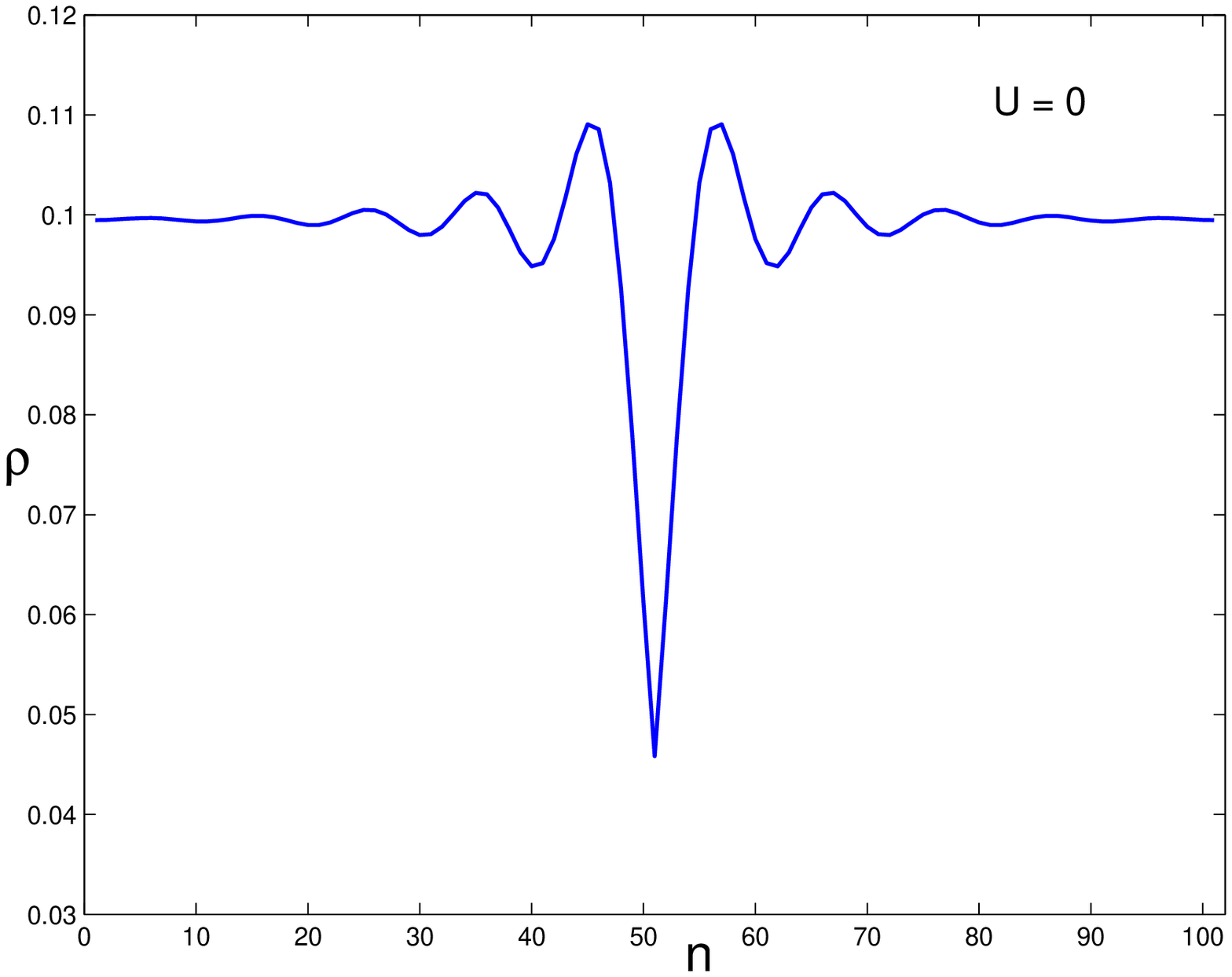,width=8.5cm}
\epsfig{figure=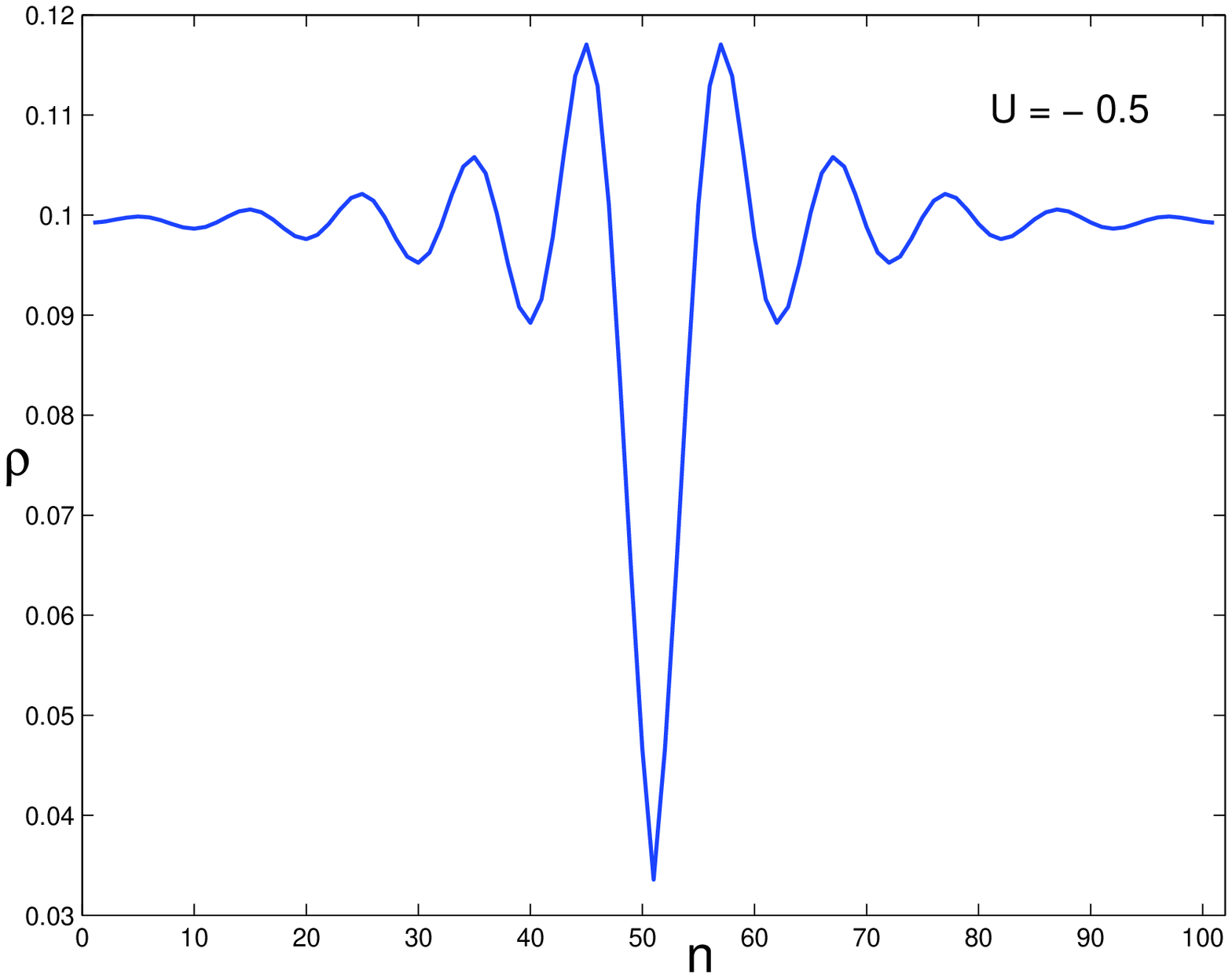,width=8.5cm}
\end{center}
\caption{(Color online) Friedel oscillations for spin-1/2 electrons for 
$U =$ 0.5, 0 and -0.5 (upper, middle and lower figures respectively) for 
$V=0.3$, $L=101$ and $\beta = 100$.}
\end{figure}

\subsection{Spin-1/2 electrons}

We now consider the spin-1/2 model with an on-site interaction between the
electrons. In the absence of interactions, the conductance at zero temperature
is independent of the wire length and is given by Eq. (\ref{g0}), with a 
factor of 2 for spin; namely, 
\beq 
G_0 ~=~ \frac{2e^2}{h} ~\frac{4 \sin^2 k_F}{4 \sin^2 k_F ~+~ V^2} ~.
\label{gvsv}
\eeq
For $k_F = \pi /10$ and $|V| = 0.03$, $G_0 = (2e^2 /h) ~0.809$.

In Fig. 7, we show the conductance as a function of the wire length $L$ for 
three different value of $\beta$, with $U=0.3$ in the upper figure and 
$U=-0.3$ in the lower figure. The trends in Fig. 7 are in accordance with 
Eq. (\ref{rb1}). Firstly, the conductance increases with the length scale if 
$U > 0$ and decreases if $U < 0$. For $U=0.3$, we have $U/ (2 \pi v_F) = 
0.0773$; we expect the HF approximation to be reasonable for such a small 
value. [Due to the RG flows of $g_1$ and $g_2$ in Eq. (\ref{rg4}), the 
conductance should start decreasing for positive $U$ and increasing for 
negative $U$ at a very large length scale of the order of $L \sim 
\exp(1/0.0773) \sim 420000$, provided that $L_T$ is also larger than $420000$. 
Such a wire length is beyond our numerical capability.]

Secondly, for fixed $\beta$, $G$ increases till $L$ reaches a value of the 
order of $L_T = 0.618 \beta$ beyond which $G$ stops changing. This happens 
because there is complete thermal decoherence once $L$ exceeds $L_T$; hence 
$G$ does not vary any more with $L$ in that regime. Thirdly, the conductance 
for two different values of $\beta$, say, $\beta_1$ and $\beta_2$ (where 
$\beta_1 < \beta_2$), start separating from each other at a value of the wire 
length $L_c$ such that $L_c /\beta_1$ is much less than $0.618$. This happens
because if the inverse temperature is $\beta_1$ or higher, there is complete 
thermal coherence, and $G$ does not depend on $\beta$ any more. Thus, for any 
temperature $\beta$, there is a range of wire lengths going from $L_c$ up to 
$L_T$ where there is partial thermal coherence, and the conductance evolves 
slowly in this range of lengths. We therefore conclude that the RG flow does 
not suddenly stop at a length scale which is the smaller of the wire length 
$L$ and the thermal coherence length $L_T$; rather, the flow continues to 
occur (but slows down) in an intermediate range of length scales.

Finally, Fig. 7 shows that even for the smallest values of $L$, the conductance
differs from the non-interacting value of $0.809$ by an appreciable amount. 
This is due to the large deviations of the density from the mean value at the
sites close to the impurity. In that region, the density does not satisfy the
$1/|n|$ form given in Eq. (\ref{rhon}). Since that form is intimately tied 
to the RG results (the sum over $1/|n|$ gives $\ln (L)$), we do not expect 
the numerically obtained value of the conductance to agree well with the RG
analysis for short wire lengths.

In Fig. 8, we show the conductance as a function of $\beta$ for three 
different value of $L$, with $U=0.3$. Once again, we see the same trends as 
in Fig. 7. Namely, for any two values of $L$, there is 
a $\beta$ where the values of $G$ start separating from each other. Then 
there is a higher value of $\beta$ where $G$ stops changing. The region 
between the two is where there is only partial thermal coherence, and $G$
varies relatively slowly in that region.

In Fig. 9, we show the conductance as a function of $V$ for three different
values of $U$, with $L=101$ and $\beta = 100$. Note that for $U=0$, the 
conductance is an even function of $V$ given by Eq. (\ref{gvsv}). But in the 
presence of interactions, $G$ is no longer precisely an even function of $V$,
particularly for large values of $V$.
This is more clearly visible in Fig. 10 which shows the conductance as a 
function of $U$ for $V=$ 1 and -1, with $L=101$ and $\beta = 100$. For any 
value of $U$, we find that the conductance deviates less from the 
non-interacting value for $V>0$ compared to $V<0$, for the same value of $|V|$.
For small values of $U$, this can be explained as follows. We saw in Figs. 3 
and 4 that the density at the site of the impurity deviates from the mean 
density of ${\bar \rho} = 1/10$ by a smaller amount for $V>0$ compared to 
$V<0$; the numerical values for $\rho_0 - 1/10$ are given by -0.046 and 0.095 
for $V=$ 0.3 and -0.3 respectively. The effective impurity strength is given by
\beq
V_{\rm eff} ~=~ V ~+~ U ~( \rho_0 ~-~ {\bar \rho}) ~,
\eeq
which equals $0.3 - 0.046 U$ and $-0.3 + 0.095 U$ for $V=0.3$ and $-0.3$ 
respectively. For $U$ small and positive, this means that the magnitude of 
the effective impurity strength and therefore the reflection probability is 
larger for $V=0.3$; hence the conductance is smaller for $V=0.3$. The opposite
statement is true if $U$ is small and negative. To conclude, the different 
values of the density deviation at the impurity site for positive and negative
values of $V$ are responsible for the asymmetry in $G$ as a function of $V$.
(This asymmetry is more prominent for the case of spinless electrons as we 
will see below).

The fact that the deviation of the conductance from the non-interacting 
value is smaller for $V>0$ than for $V<0$ is clearly 
a result of the bound state which raises the density
at the impurity site for $V<0$; hence this is a short distance effect.
[The contribution of the scattering states to the reflection probability
$|r|^2$ is an even function of $V$ as one can see from Eq. (\ref{rb1}).
The RG equations which are derived from continuum theories take into account 
only long distance effects; these come from the scattering states only.]

In Fig. 11, we show the Friedel oscillations for three different values of $U$
with $V=0.3$, $L=101$ and $\beta = 100$. We see that a repulsive interaction 
($U>0$) suppresses the Friedel oscillations, while an attractive interaction
($U<0$) with the same magnitude enhances the oscillations by a much larger
amount. This is because attractive interactions tend to lead to the formation
of a charge density wave; if a charge density wave is already present (due to
the impurity), attractive interactions enhance it. Since the scattering from 
the Friedel oscillations renormalize the scattering from the impurity, this 
difference in the magnitude of the oscillations helps to explain why the 
deviation of the conductance $G$ from the non-interacting value is larger 
for $U<0$ compared to $U>0$, as can be seen in Fig. 7.

\begin{figure}[htb] 
\begin{center}
\epsfig{figure=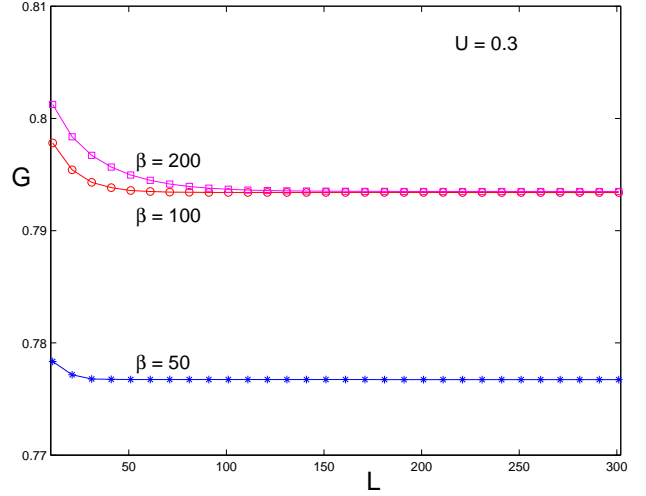,width=8.5cm}
\end{center}
\caption{(Color online) Conductance versus wire length for interacting 
spinless electrons for three different values of $\beta$, for $V=0.3$ and 
$U=0.3$.}
\end{figure}

\begin{figure}[h!]
\begin{center}
\epsfig{figure=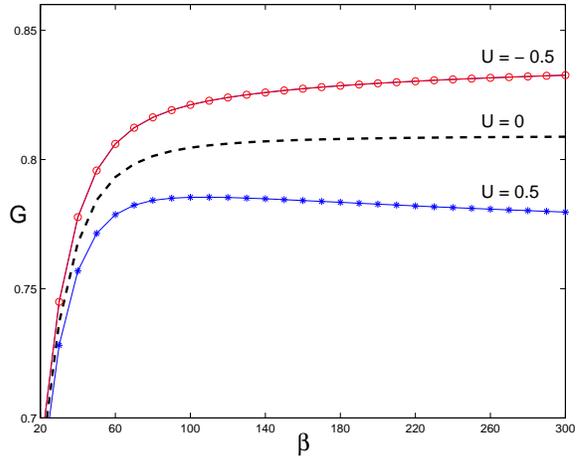,width=8.5cm}
\end{center}
\caption{(Color online) Conductance versus $\beta$ for interacting spinless 
electrons for three different values of $U$, for a wire with 201 sites and 
$V=0.3$.}
\end{figure}

\begin{figure}[htb]
\begin{center}
\epsfig{figure=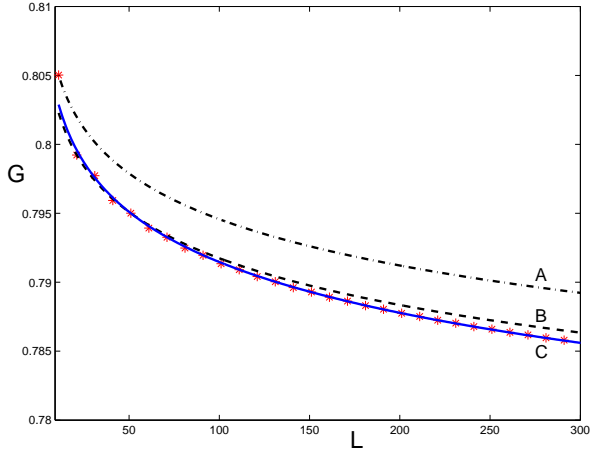,width=8.6cm}
\end{center}
\caption{(Color online) Conductance $G$ versus $L$ for interacting spinless 
electrons compared with the RG expression in Eq. (\ref{solvrg}), for $V = 0.3$,
$U=0.3$, and $\beta = \infty$. Line A (dash dot line) corresponds to $d=11, ~
|t(d)|^2 = 0.805$ and $\alpha = 0.0148$. Line B (dashed line) corresponds 
to $d=51, ~|t(d)|^2 = 0.795$ and $\alpha = 0.0148$. Line C (solid line) 
corresponds to $d=51, ~|t(d)|^2 = 0.795$ and $\alpha = 0.016$. The asterisks 
show the numerical results.}
\end{figure}

\begin{figure}[h!]
\begin{center}
\epsfig{figure=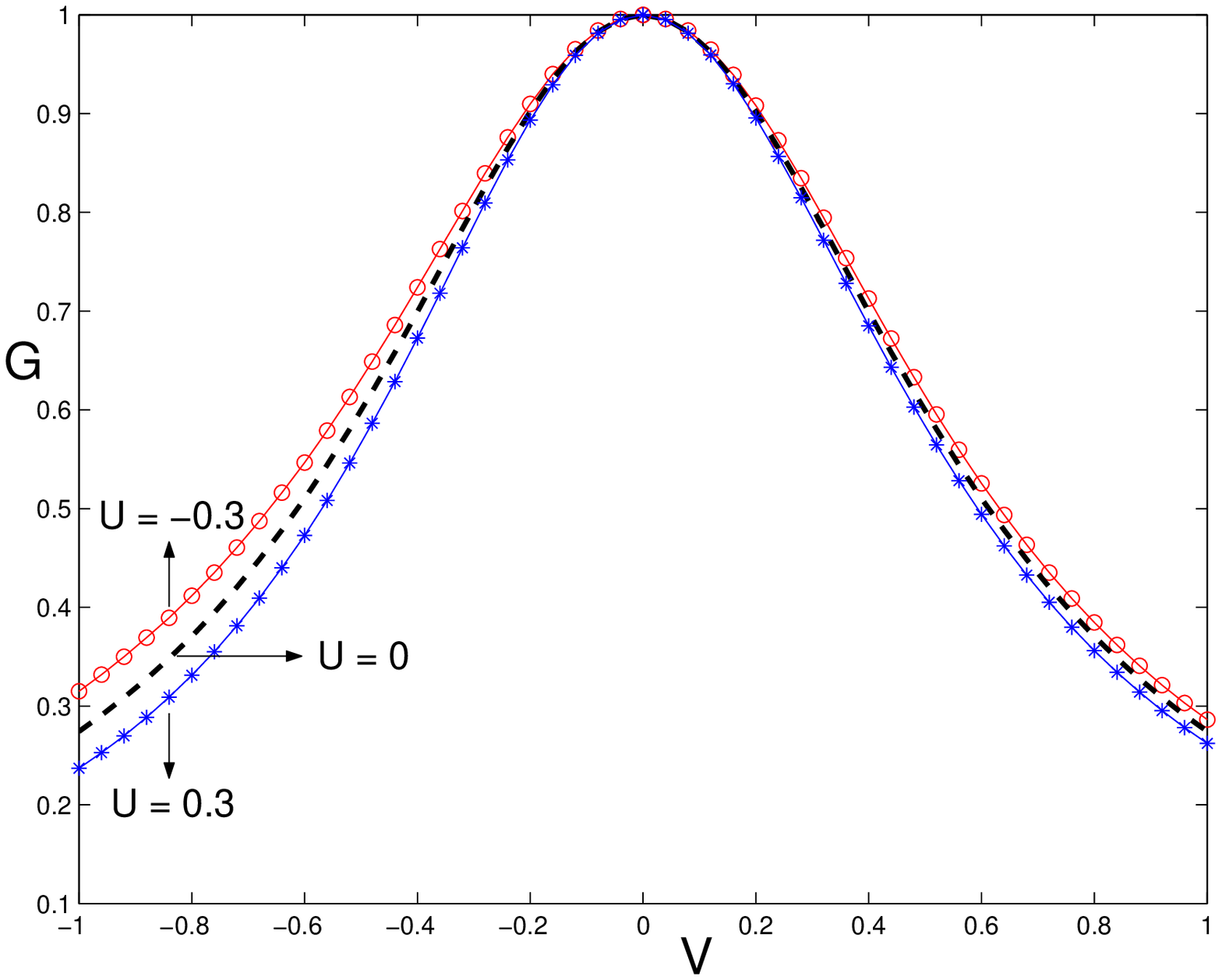,width=8.5cm}
\end{center}
\caption{(Color online) Conductance $G$ versus $V$ for interacting spinless 
electrons for $U$ = 0.3, 0 and -0.3, with $L=101$ and $\beta = 100$.}
\end{figure}

\begin{figure}[htb]
\begin{center}
\epsfig{figure=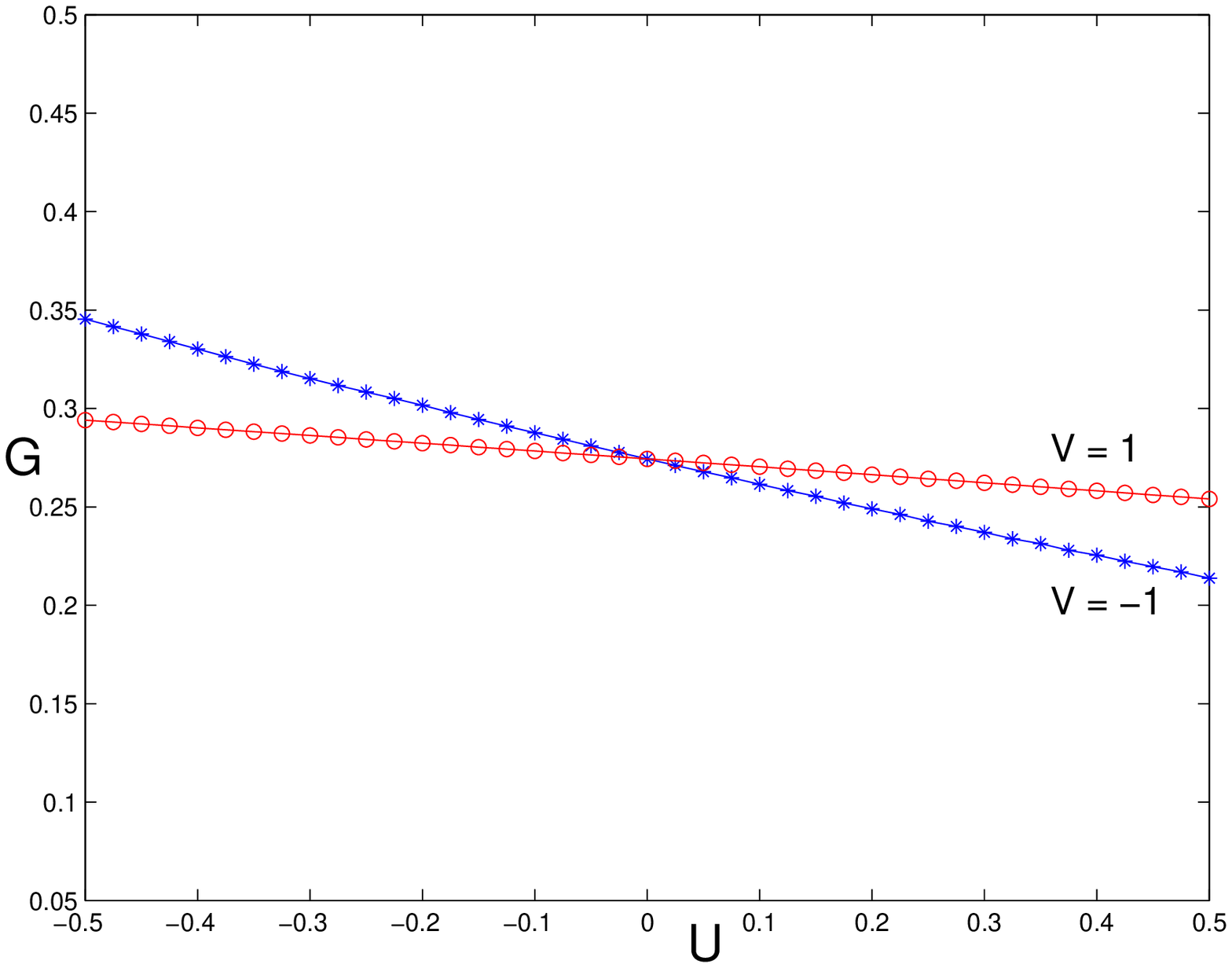,width=8.5cm}
\end{center}
\caption{(Color online) Conductance $G$ versus $U$ for interacting spinless 
electrons for $V$ = 1 and -1, with $L=101$ and $\beta = 100$.}
\end{figure}

For spin-1/2 electrons, we can do a self-consistent HF calculation in two ways,
restricted and unrestricted. In a restricted HF calculation, the site densities
of spin-up and spin-down electrons are taken to be equal at all stages. 
In a unrestricted HF calculation, spin-up and spin-down electrons are allowed 
to have different densities. We have done both kinds of calculations. For the 
range of the interaction $-0.5 \le U \le 0.5$, we find that they give the same
results; the spin-up and spin-down densities converge to the same values even 
if one begins with different initial values for them. Thus we do not find any 
spin density wave within our range of parameters.

\subsection{Spinless electrons} 

We now consider a model of spinless electrons with nearest neighbor 
interactions between the electrons. Comparing the forms of Eqs. (\ref{rb1}) 
and (\ref{rb3}), we see that the behaviors of the spin-1/2 model for $U>0$ 
and $U<0$ should be similar to that of the spinless model for $U<0$ and $U>0$ 
respectively. In Fig. 12, we show the conductance as a function of the wire 
length $L$ for three different value of $\beta$, with $V=0.3$ and $U=0.3$. 
The trends are in agreement with Eq. (\ref{rb1}), with $\alpha = U (1 - \cos 
(2k_F))/ (2 \pi v_F) = 0.0148$. For fixed $\beta$, $G$ decreases till $L$ 
reaches a value of the order of $L_T = 0.618 \beta$.

In Fig. 13, we show the conductance as a function of $\beta$ for three 
different value of $U$, with $V=0.3$ and $L=201$. As we saw earlier in 
Fig. 6, $G$ has a significant dependence on $\beta$ for small values of
$\beta$ even for non-interacting electrons. It is only when we go to
large values of $\beta$ that we see the trend expected from the RG 
equations for interacting electrons; namely, as $\beta$ increases, the 
conductance decreases if $U>0$ and increases if $U<0$.

Since $G$ has an appreciable dependence on $\beta$ even for non-interacting 
electrons, it is easier to consider the dependence of $G$ on the wire 
length $L$ at zero temperature in order to see how well the numerical
results compare with the RG expression given in Eq. (\ref{solvrg}). That
equation has two parameters, namely, the interaction parameter $\alpha$
and the short distance scale $d$ (with its corresponding transmission 
probability $|t(d)|^2$). To begin, let us choose $\alpha = 0.0148$ as given 
by the analytical expression in Eq. (\ref{rb3}) in terms of $U$ and $v_F$.
In Fig. 14, we show a comparison between the expression in (\ref{solvrg}) for 
two values of $d$, namely, $d=11$ with $|t(d)|^2 = 0.805$ (the dash dot line 
A), and $d=51$ with $|t(d)|^2 = 0.795$ (the dashed line B), with our numerical
results shown by astersisks. (The values of $|t(d)|^2$ for $d=$ 11 and 51 have
themselves been obtained by our numerical calculations). We see that the RG 
expression with $d=51$ fits the numerical results better than the expression 
with $d=11$.
This difference between the lines A and B shows that the interactions change 
the conductance by an appreciable amount between $d=11$ and $d=51$, and this 
change is not captured accurately by the expression in Eq. (\ref{solvrg}).
[The conductance at short distances is affected substantially by quantities
such as the large value of $\rho - {\bar \rho}$ at the site of the impurity. 
The RG analysis, on which (\ref{solvrg}) is based, does not take such effects
into account.] Since the wavelength of the Friedel oscillations is given by 
$\lambda = 10$, we conclude that the RG results agree reasonably well with 
the numerical results only if we begin the integration of the RG flows from a 
distance which is significantly larger than $\lambda$. However, we observe that
even line B corresponding to $d=51$ starts deviating from the numerical results
at large length scales. We have therefore tried varying the parameter $\alpha$ 
also, keeping $d$ and $|t(d)|^2$ fixed at $51$ and $0.795$ respectively. We 
find that $\alpha = 0.016$ (the solid line C in Fig. 14) gives an excellent 
fit to the numerical results.

In Fig. 15, we show the conductance as a function of $V$ for three different
values of $U$, with $L=101$ and $\beta = 100$. We see that $G$ is not an even 
function of $V$ in the presence of interactions.
In Fig. 16, we show the conductance as a function of $U$ for $V=$ 1 and -1, 
with $L=101$ and $\beta = 100$. For any value of $U$, we again find that 
the conductance deviates less from the non-interacting value for $V=1$ 
compared to $V=-1$. The explanation for this is similar to that for the 
similar phenomenon in the spin-1/2 model, except that the roles of $U$ and 
$-U$ are interchanged in the two models. The larger change in the conductance 
for $V=-1$ is again due to the presence of a bound state.

\section{Discussion}

We have used the NEGF formalism to study the dependence of the 
conductance of a quantum wire with both spin-1/2 and spinless electrons on 
various parameters such as the wire length, temperature, impurity potential, 
and the strength of the interactions between the electrons. The advantage of
the NEGF formalism is that it can be used to compute the density and
conductance at any length scale. At large length scales, our numerical 
results agree with those obtained by an RG analysis of continuum theories. 
We find that the trends of the RG results can be understood using the Born 
approximation for scattering from a weak impurity and from the density 
oscillations produced by the impurity. 

Our numerical results differ in detail in two ways from those obtained 
analytically from the RG equations. Firstly, the dependence of $G$ on the wire
length and temperature fits the expression in Eq. (\ref{solvrg}) only if we 
take the starting point of the RG equation to be significantly larger than 
the short distance scale $\lambda$, and we allow $\alpha$ to be a little 
different from its analytically obtained value. Secondly, the conductance is 
not an even function of the impurity potential $V$ (as one expects from the 
RG analysis); this is due to the existence of a bound state for an attractive 
impurity. These differences between the numerical results and 
the results based on the RG analysis seem to be due to effects at short 
distances, where the density deviates significantly from the mean density. 

Before ending, we would like to briefly compare our work with that of Refs. 
\cite{enss,ander,barnabe}. Using the functional RG technique, the 
authors of those papers have shown that there is excellent agreement at very 
large length scales between the numerical results and the asymptotic scaling 
forms given by the RG equations derived from continuum theories.
We have not been able to go up to such large length scales using the NEGF
formalism. However, even at the length scales studied by us, our numerical 
results agree quite well with the RG equations if we start at a short 
distance scale which is about 5 times larger than $\lambda$. 

As mentioned towards the beginning of Sec. V, the length scales $L$ and $L_T$ 
that we have considered are comparable to those studied experimentally. Our 
observations about the short distance effects may therefore have implications
for fitting experimental data to expressions obtained by an RG analysis.

\vskip .5 true cm
\centerline{\bf Acknowledgments}
\vskip .5 true cm

We thank Supriyo Datta, Avik Ghosh and V. Ravi Chandra for many stimulating 
discussions. We thank the Department of Science and Technology, India for 
financial support under projects SR/FST/PSI-022/2000 and SP/S2/M-11/2000.

\end{document}